\documentclass[12pt]{article}
\usepackage[margin=1in]{geometry}
\usepackage{authblk}                 % per-author affiliation markers
\usepackage[numbers,sort&compress]{natbib}
\usepackage{graphicx}
\usepackage{booktabs}
\usepackage{multirow}
\usepackage{enumitem}
\usepackage{url}
\usepackage{amsmath,amssymb}
\usepackage[hidelinks]{hyperref}

\begin{document}

%\title{GenAI-Assisted Data Comics for Education: Evaluating Effectiveness, Benefits and Limitations}
% \title{Data Comics for Education: Insight Communication, Visualisation Literacy, and the Role of Generative AI}
\title{Data Comics for Education: Evaluating Effectiveness, Benefits, and the Ethics of AI-Assisted Creation}

\author[1]{Zirui Shan}
\author[2]{Vanessa Echeverria}
\author[3]{Yuheng Li}
\author[1]{Yi-Shan Tsai}
\author[1]{Roberto Martinez-Maldonado\thanks{Corresponding author: Roberto.MartinezMaldonado@monash.edu}}

\affil[1]{Faculty of Information Technology, Monash University, Melbourne, Australia}
\affil[2]{School of Computing Technologies, RMIT University, Melbourne, Australia}
\affil[3]{Department of Applied Social Sciences, The Hong Kong Polytechnic University, Hong Kong, China}
% \author[1]{Blinded for Review}

% \affil[1]{Blinded for Review}

\date{}

\maketitle

\begin{abstract}
In today's data-driven world, students often struggle with interpreting visualisations due to limited visualisation literacy. Data comics have emerged as a promising medium to enhance engagement and understanding, but their educational value has seen little empirical examination, partly due to the effort required to create them. Recent advances in Generative AI (GenAI) offer a scalable solution to this challenge. We conducted a within-subjects study with 60 university students, comparing conventional visualisations with data comics, created with assistance from GenAI tools, across information retrieval and comprehension tasks. Students consistently performed better with data comics, particularly in insight comprehension tasks, independent of prior visualisation literacy. Students also commented data comics as more engaging and easier to understand, though concerns were raised about GenAI-driven misinformation and ownership. Our findings highlight the potential of data comics as a potentially effective tool for data communication in education, while underscoring the need to address ethical concerns related to AI-assisted creation.
\end{abstract}


%\linenumbers

\section{Introduction}\label{sec1}
In recent years, technological advancements have generated vast amounts of data, driving innovations in analytics, which now underpin decision-making processes across industries \citep{chen2014big, sagiroglu2013big}. Higher education, therefore, faces the challenge of preparing students from diverse disciplines to develop foundational skills in interpreting and using data for reasoning, problem-solving and critical thinking as part of their future professional practice \citep{maybee2015data, qiao2024understanding, pratama2020data}. 

To support these goals, educators have adopted a wide range of data visualisation approaches as a means for translating complex data into accessible visual formats \citep{milesi2024s, donohoe2020data}. These visualisations, ranging from simple line charts to intricate alluvial diagrams, aim to communicate data-driven insights effectively \citep{shen2019information}. 
However, students' capacity to derive meaning from these visualisations, including identifying key data points and interpreting the insights they convey, is not solely a function of their exposure to them. It is shaped both by their level of visualisation literacy -- the skills required to interpret and reason with visual data \citep{roberts2017explanatory, donohoe2020data} -- and by the design choices that determine how information is encoded and presented in more intuitive ways \citep{tufte2001visual}.
Indeed, research suggests that students, including those in STEM-related disciplines (Science, Technology, Engineering, Mathematics) \citep{maltese2015data}, often struggle to interpret data visualisations presented during class \citep{bodily2017trends, fernandez2022beyond}.

One response to these educational challenges has been a trend toward adopting data storytelling, which integrates narrative structures with data visualisations to guide attention, highlight trends, and provide context for interpretation \citep{knaflic2015storytelling,kosara2013storytelling,dykes2015data}, presenting a promising design space for educational use \citep{echeverria2018driving,fernandez2022beyond,Fernandeznieto2025steditor}. 
Within this genre, data comics have emerged as a promising format that adapts the familiar structure of comics, such as panels, sequencing, characters, and narratives, to communicate data-driven stories \citep{bach2017emerging,bach2018design,wang2019teaching,wang2020cheatsheet}. This approach has been suggested as particularly supportive for novice students, while also fostering the gradual development of visualisation literacy 
\citep{wang2020cheatsheet,boucher2023educational}.
%
%Suggested as a promising method for simplifying the communication of complex data insights, data comics promise to support novice students while fostering the development of visualisation skills \citep{wang2020cheatsheet, boucher2023educational}. By integrating visual and textual storytelling techniques, data comics may have the advantage of contextualising information, employing sequential layouts to gradually reveal data-driven narratives \citep{bach2018design, wang2019teaching, wang2020cheatsheet}. The growing exploration of data comics, from K-12 to higher education \citep{vacca2022happen, matuk2022data, hasan2022playing, tes2023data}, as tools to enhance the accessibility of embedded information and encourage active exploration of data insights, highlights their potential as effective educational tools in fields that require students to understand and engage with data.
%
%Data comics have been suggested as a prospective approach for simplifying the communication of complex insights, supporting novice students while fostering the development of visualisation skills \citep{wang2020cheatsheet,boucher2023educational}. By integrating visual and textual storytelling techniques, data comics contextualise information using sequential layouts to gradually reveal data-driven narratives \citep{bach2018design,wang2019teaching,wang2020cheatsheet}. 
For education, data comics opens a unique design space to simplify the communication of complex insights \citep{bach2018design}. Evidence of their growing use across K–12 \citep{matuk2022data,tes2023data,vacca2022happen,hasan2022playing} and higher education \citep{milesi2024s,Boucher2025instructional} highlights their potential as educational tools that can enhance students' engagement \citep{matuk2022data}, and encourage active exploration of data insights \citep{hasan2022playing,Boucher2025instructional}.

However, although some authors have conceptually suggested that data comics are novice-friendly and potentially beneficial in educational settings \cite[e.g.,][]{lc2022designing,boucher2023educational}, empirical evidence on their actual effectiveness remains limited, especially in direct comparisons to conventional visualisations. A significant gap exists in understanding students' perceptions of data comics and whether they can effectively interpret them to grasp data insights.
One of the main obstacles that has long hindered such investigations is the time-consuming and labour-intensive creation process, which often relies on manual design and crafting \citep{milesi2024s,Kangtoonnote2021}, and requires expertise and skills across multiple domains such as data visualisation, art, storytelling, and pedagogy \citep{boucher2023educational,Boucher2025instructional}. This production bottleneck constrains opportunities to systematically evaluate their educational potential.

Yet, recent advancements in generative artificial intelligence (GenAI), particularly text-to-image GenAI tools (e.g., \textit{Midjourney}, \textit{DALL-E 3}), offer new opportunities to address this production bottleneck. Researchers and practitioners can now use such tools to accelerate the creation of visual content and narratives for building data comics \citep{huang2023future,milesi2024s,zhang2023text}. Yet, despite the potential benefits, using GenAI for educational purposes raises ethical concerns related to critical aspects such as misinformation, bias, ownership and transparency \citep{yan2024vizchat,farrelly2023generative}. Thus, understanding both the educational potential and the ethical risks, before deployment in authentic settings, is essential for exploring how GenAI and data comics can be integrated to support learning tasks.

To this end, we conducted a comparative study with 60 student participants to examine: 
i) the extent to which data comics, co-created with GenAI assistance, \textbf{effectively} communicate data insights compared to conventional visualisations for \textbf{information retrieval} and \textbf{comprehension} tasks; 
ii) the potential mediating role of students' \textbf{visualisation literacy}; and
iii) students' reflections on the \textbf{benefits and limitations} of these data comics, as well as \textbf{ethical concerns} regarding their AI-assisted creation for educational purposes. 
Effectiveness was assessed through students' accuracy in responding to questions about the data represented through visualisations or data comics, aligning with prior comparative studies of data visualisation \citep{shao2024data,milesi2025piecing} (i). The Mini-VLAT instrument was used to measure students' visualisation literacy \citep{pandey2023mini} (ii), while thematic analysis captured perceptions of benefits, limitations, and ethical implications (iii).  To the best of our knowledge, this is the first study to empirically investigate the effectiveness of communicating data insights to students through data comics co-created with GenAI technologies.

\section{Background and Related Works}
\subsection{Foundations of Data Storytelling in Education} %and Data Comics in Educational Contexts}

%Conventional data visualisation employs computational techniques to transform raw data into visual forms, enabling viewers to explore, uncover, and interpret information \citep{Hinterberger2009}. Depending on the data type (e.g., continuous, ordinal), various conventional visualisation techniques, such as line charts, bar charts, and boxplots, are commonly used to facilitate interpretation \citep{roberts2017explanatory}. While exploratory visualisations allow viewers to visually inspect data and potentially identify insights \citep{hicks2009perceptual}, fully understanding the practical implications of the data often requires individuals to possess the skills necessary to interact effectively with these visualisations \citep{martinez2020data, srinivasan2018augmenting, lee2016vlat, boy2014principled}. 

Conventional data visualisation transforms raw data into visual forms that enable viewers to explore, uncover, and interpret information \citep{Hinterberger2009}. Various techniques such as line charts, bar charts and boxplots are commonly used to facilitate interpretation \citep{roberts2017explanatory}. However, fully understanding the insights extracted from data often requires individual's ability to interact effectively with these visualisations \citep{martinez2020data, srinivasan2018augmenting, lee2016vlat, boy2014principled}. 

%In educational contexts, the ability to interpret and make sense of the insights derived from data visualisations often requires the student’s ability to connect to their own prior knowledge and experiences \citep{honebein1996seven,abderrahim2021theoretical}. This is oftentimes associated with \textit{constructivism theory}, which posits that learning is an active process where individuals build understanding by integrating new information with their existing cognitive frameworks and experiences \citep{cashman2008power,phillips1995good,abderrahim2021theoretical}.However, students bring differing levels of background in data and visualisation, which means their ability to extract insights from standard visual forms can vary widely. Prior studies show that many struggle to interpret visualisations in learning dashboards \citep{bodily2017trends, fernandez2022beyond}, and even STEM undergraduates also find it challenging to understand simple data visualisations presented in class slides \citep{maltese2015data}. Thus, limited visualisation literacy can prevent students from extracting meaningful insights from conventional data visualisations \citep{borner2016investigating, maltese2015data}. This ineffectiveness may heighten when visualisations lack the contextual and cognitive support needed for students to connect new information with their prior knowledge \citep{talebi2015john}. To address this, data storytelling has been proposed as a novice-friendly, constructively aligned approach to communicating insights \citep{knaflic2015storytelling}.

In educational contexts, interpreting these data insights often expects students' ability to connect new information with prior knowledge and experiences \citep{honebein1996seven,abderrahim2021theoretical}. This aligns with \textit{constructivism theory}, which views learning as an active process of integrating new information into existing cognitive frameworks \citep{cashman2008power,phillips1995good,abderrahim2021theoretical}. Yet, students bring varying level of background in data visualisation, making interpretation challenging. Prior studies show difficulties in interpreting visualisations in learning dashboards \citep{bodily2017trends, fernandez2022beyond} and even simple classroom charts among STEM undergraduates \citep{maltese2015data}. Thus, limited visualisation literacy can hinder insight extraction especially when visualisations lack the contextual and cognitive support \citep{borner2016investigating, maltese2015data, talebi2015john}. To address these challenges, data storytelling has been proposed as a novice-friendly, constructively aligned approach to communicating insights \citep{knaflic2015storytelling}.

%Data storytelling is a design-oriented approach that combines narrative elements, such as annotations, descriptions, and sequencing, with visualisation design features, including colour, arrows, and clear titles, to highlight the most relevant data points and direct attention to critical insights \citep{dykes2015data}. A central design premise of data storytelling in visualisation is the intentional emphasis on data points that are most relevant to the story or message, while de-emphasising less critical information \citep{knaflic2015storytelling}. Through purposeful design choices, visual emphasis and textual narratives work together to enhance comprehension of insights identified during data analysis and to foreground key information that supports the overall findings \citep{shao2024data}.

Data storytelling combines narrative elements (e.g., annotations, descriptions, sequencing) with visualisation design features (e.g., colour, arrows, clear titles) to guide attention towards critical insights \citep{dykes2015data}. By emphasising on the most relevant data points and providing contextual annotations intentionally, data storytelling can make insights more interpretable and meaningful \citep{knaflic2015storytelling, shao2024data}.

%In education, some studies have examined the impact of data storytelling on educators' interpretation and use of data insights. For instance, in a study, educators were found to derive more actionable insights from visual analytics when storytelling elements facilitated comprehension \citep{echeverria2018exploratory}. Eye-tracking methods have also demonstrated how annotations embedded in data visualisations can direct teachers’ attention to specific points relevant to the story being conveyed \citep{echeverria2018driving}. Further, inviting educators to co-edit data visualisations and enhance them with narratives has shed light on the perceived benefits and concerns surrounding data storytelling \citep{milesi2024qualitative}. While these studies highlight the potential effectiveness of data storytelling for educators, empirical evidence regarding its impact on students remains scarce. One exception examined the use of annotated charts with healthcare students and found that storytelling elements supported learners in gaining insights on their own practice from visualised data \citep{martinez2020data}. Collectively, this body of work suggests that exploring specific genres of data storytelling, such as data comics, may be timely for supporting novice students \citep{boucher2023educational, bodily2017trends, fernandez2022beyond}.

Studies in education has primarily examined the impact of data storytelling on educators. Storytelling elements have been shown to facilitate educators' comprehension of visual analytics \citep{echeverria2018exploratory}, direct attention to relevant insights \citep{echeverria2018driving}, and support reflection \citep{milesi2024qualitative} through narrative-embedded visualisation. While annotated charts have been found to support healthcare students in gaining insights on their own practice, empirical evidence regarding its impact on students remains scarce \citep{martinez2020data}. Collectively, this body of work suggests that exploring specific genres of data storytelling, such as data comics, may be timely for supporting novice students \citep{boucher2023educational, bodily2017trends, fernandez2022beyond}.

\subsection{Data Comics for Student Learning}

%The \textit{comic strip} is one of the seven distinct data storytelling genres proposed by \citet{segel2010narrative} to communicate data insights. As a rather novel branch of data storytelling inspired by the comic strip format, \textit{data comics} commonly utilise similar design principles (e.g., frame, image, word) as in traditional comics \citep{tobita2011comic} and integrate the narrative elements (e.g., textual and pictorial content) drawn from data visualisations into panel layouts to convey insights from data \citep{wang2019teaching,bach2017emerging}. The adoption of data comics for communicating data-driven narratives aims to maximise the clarity of the communicated insights to a broad audience, including both experts and non-experts \citep{bach2018design,wang2019teaching,liang2024data}. 

As a rather novel branch of data storytelling inspired by comic strips \citep{segel2010narrative}, \textit{Data comics} utilise similar design principles (e.g., frame, image, word) as in traditional comics \citep{tobita2011comic} and integrating narrative elements (e.g., textual and pictorial content) into panel layouts to convey insights from data \citep{wang2019teaching,bach2017emerging}. The adoption of data comics for communicating data-driven narratives aims to maximise the clarity of the communicated insights to both experts and non-experts \citep{bach2018design,wang2019teaching,liang2024data}.

For this purpose, \citet{bach2018design} delineated two fundamental dimensions (i.e., content relation and layout) to form several design patterns that guide how these dimensions can be organised to facilitate insight communication coherently. Some researchers \citep[e.g.,][]{liang2024data,milesi2024s} have applied these patterns to communicate convoluted data insights effectively and address the skill gap that non-experts, such as students, may have to understand conventional visualisations \citep{maltese2015data, bodily2017trends, fernandez2022beyond}. For instance, \citet{liang2024data} used data comics to present knowledge graphs of endangered Indigenous Australian languages, finding that narrative elements, presented with detailed information, scaffold easier understanding for both educators and students. Similarly, \citet{milesi2024s} transformed conventional visualisations from educational simulations into data comics to support students' engagement and reflection.

%However, the effectiveness of data comics in conveying data insights to students compared to conventional visualisations remains underexplored. The closest study is by \citet{wang2019comparing}, who compared data comics with infographics (visual representations combining imagery, visualisations, and minimal text) and found comics to be more enjoyable and easier to follow, leading to improved comprehension and recall. A related study by \citet{gomez2023personal} examined personal data comics with non-student participants and reported higher engagement compared to annotated visualisations. However, while data comics showed advantages in efficiency, accuracy, and recall, these differences were not statistically significant. Moreover, the study was limited by its single comparison set and its lack of attention to participants' visualisation literacy.

However, the effectiveness of data comics in conveying data insights to students compared to conventional visualisations remains underexplored. The closest study is by \citet{wang2019comparing}, who compared data comics with infographics and found comics to be more enjoyable and easier to follow, leading to improved comprehension and recall. Likewise, \citet{gomez2023personal} reported higher engagement with personal data comics than annotated visualisations, while differences in efficiency, accuracy, and recall were not statistically significant. Moreover, the study was limited by its single comparison set and lack of attention to participants' visualisation literacy.

%These limitations underscore the need for further research into the role of data comics in student learning. In particular, it remains to be examined: (1) whether data comics are more, less, or equally effective than conventional visualisations for communicating insights; (2) how students with different levels of visualisation literacy navigate data comics as a form of data storytelling; and (3) how students perceive the benefits, limitations, and overall effectiveness of data comics for insight communication.

Consequently, it remains to be examined: (1) whether data comics are more, less, or equally effective than conventional visualisations for communicating insights; (2) how students with different levels of visualisation literacy navigate data comics; and (3) how students perceive the benefits, limitations, and effectiveness of data comics for insight communication. Addressing these questions has traditionally been constrained by the time and effort required to create data comics \citep{milesi2024s,Kangtoonnote2021}. Recent advances in AI-assisted content generation, however, offer exciting opportunities to accelerate data comic creation and allow the wider adoption in educational research and practice \citep{bach2018design, milesi2024qualitative,milesi2024s}. This opens the door to systematically exploring the educational potential of data comics at scale.

%Yet, creating data comics for such research purposes is time-consuming and labour-intensive \citep{milesi2024s,Kangtoonnote2021}. the advancements in AI technologies offer exciting opportunities to support the creation of data comics, enabling researchers to generate them without the artistic skills traditionally required to draw comics from scratch \citep{bach2018design}, and allowing designers of educational interventions and tools to scale their use in practice \citep{milesi2024qualitative,milesi2024s}. This opens the door to systematically exploring the research gaps outlined above by leveraging AI technologies to support data comics creation.

\subsection{AI-Assisted Creation of Data Storytelling and Data Comics}

%Recent advancements in GenAI are transforming the ways in which people engage with data, particularly in retrieving key information and comprehending the insights contained within datasets \citep{huang2023future}. Continuous progress in GenAI has enabled multimodal content generation, ranging from text-to-text (T2T) \citep{baidoo2023education} to text-to-image (T2I) \citep{zhang2023text} and text-to-video (T2V) \citep{khachatryan2023text2video}. These capabilities can substantially reduce the time and effort traditionally required to produce media, thereby scaling practices that once demanded extensive manual work. Within the context of data storytelling, GenAI introduces new opportunities for the automated creation of data-driven narratives \citep{Li2024where}.

Recent advancements in GenAI are transforming the ways in which people engage with data \citep{huang2023future}, particularly in enabling multimodal content generation, ranging from text-to-text (T2T) \citep{baidoo2023education} to text-to-image (T2I) \citep{zhang2023text} and text-to-video (T2V) \citep{khachatryan2023text2video}. These capabilities substantially reduce the time and effort required to produce media and introduce new opportunities for the automated creation of data-driven narratives \citep{Li2024where}.

%Early approaches to automated data storytelling, such as Calliope \citep{Shi2020calliope}, relied on logic-oriented search algorithms to identify ``data facts'' and sequence them into coherent structures. However, the field is rapidly shifting towards generative architectures that offer greater stylistic flexibility and narrative depth. For instance, \citet{Shen2024from} introduced the ``Data Director'', a multi-agent Large Language Model (LLM) framework that decomposes the storytelling process into specialised roles, such as data analysis, scriptwriting, and visual direction, to automate the creation of animated data narratives. Building on this multi-agent paradigm, \citet{Venkatraman2025collab} recently proposed ``CollabStory'', a framework that facilitates collaboration between multiple LLMs to co-author complex narratives, addressing issues of coherence that often plague single-model generation.

Early automated data storytelling systems, such as Calliope \citep{Shi2020calliope}, relied on rule-based approaches to identify and sequence ``data facts'' into coherent structures. More recently, the field has rapidly shifted towards Large Language Model (LLM)-based multi-agent frameworks (e.g., ``Data Director'') that decomposed data analysis, narrative generation, and visual design, enabling more flexible and sophisticated storytelling workflows \citep{Shen2024from,Venkatraman2025collab}. 

%In the specific domain of comics, \citet{Chen2024collaborative} proposed a collaborative generation system that integrates visual narrative theories and ``comic authoring idioms''. By encoding narrative grammars (e.g., transition types between panels) as constraints, their system allows for the generation of comics that maintain structural coherence while leveraging the creative capabilities of diffusion models. Beyond structural automation, researchers are exploring how GenAI can translate abstract technical data into accessible visual metaphors. \citet{Heidrich2023visualizing} demonstrated a pipeline for visualising software source code as comics using Stable Diffusion. Their approach adopts a chained prompting strategy where an LLM first interprets the code's logic and then generates metaphorical descriptions (e.g., representing a function as a postal worker) to drive image synthesis. To address the challenge of ``hallucination'' in these generative pipelines where models might misrepresent data values, \citet{Shi2024constraint} proposed a taxonomy of constraints and the use of Domain-Specific Languages (DSLs) to ensure data fidelity is preserved during the transition from raw data to visual narrative.

In the specific domain of comics, \citet{Chen2024collaborative} proposed a collaborative generation system that integrates visual narrative theories and ``comic authoring idioms''. By encoding narrative grammars (e.g., transition types between panels) as constraints, their system generates comics that maintain structural coherence while leveraging the creative capabilities of diffusion models. Beyond structural automation, researchers are exploring leveraging AI to translate technical data into accessible visual metaphors. \citet{Heidrich2023visualizing} developed a pipeline that uses an LLM to interpret software source code and generate metaphorical descriptions that guide comic creation through Stable Diffusion. To address concerns about ``hallucination'' and data misrepresentation, \citet{Shi2024constraint} proposed a taxonomy of constraints and Domain-Specific Languages (DSLs) to preserve data fidelity throughout the visual generation process.

%In educational contexts, these AI-enhanced formats are beginning to be evaluated for their pedagogical effectiveness. \citet{milesi2024s} employed GenAI tools to transform students' learning analytics data into personalised ``hero's journey'' comic strips. Their study found that while the narrative format significantly increased student engagement and emotional connection to the data, there was a tension regarding the perceived ``professionalism'' of the comic medium. While not focusing directly on data narratives or comics, \citet{Cui2025promises} provided a critical evaluation of the capabilities of GenAI tools in generating visualisation assessment items. Their study highlighted both the promises and pitfalls of using LLMs for such a purpose, emphasising that while GenAI can scale the creation of educational materials, rigorous validation is required to prevent subtle visual or factual errors.

In educational contexts, these AI-enhanced formats are beginning to be evaluated for their pedagogical effectiveness. \citet{milesi2024s} employed GenAI tools to transform learning analytics data into personalised comic strips and found significantly increased student engagement and emotional connection to the data. However, concerns remained regarding the perceived ``professionalism'' of the comic medium. More broadly, evaluations of GenAI-generated educational materials have highlighted both the scalability benefits and the need for rigorous validation to prevent subtle visual or factual errors \citep{Cui2025promises}.

Collectively, these studies illustrate the significant potential of AI technologies, especially GenAI, in facilitating the creation of data storytelling artefacts, while serving as a reminder that their successful integration into education requires careful consideration of its appropriateness to meet the pedagogical intent. 

\subsection{Ethical Concerns of using Generative AI in Education}\label{ethical_concerns}
While there is considerable enthusiasm about the opportunities that GenAI offers for supporting new forms of interaction with data in educational contexts, its integration is shaped by valid and pressing ethical concerns raised by educational stakeholders \citep{alasadi2023generative, michel2023challenges}. A primary challenge is the risk of dissemination of \textit{misinformation}. While GenAI holds the potential of enhancing personalised learning through tailored feedback and adaptive interventions, it also risks generating content that appears authoritative but lacks factual accuracy, potentially misleading both students and educators \citep{yan2024vizchat}. Additionally, concerns about \textit{bias and fairness} can emerge when GenAI models trained on biased data perpetuate stereotypes, as noted by \citet{farrelly2023generative}. This can compromise educational fairness, especially for diverse student groups. Moreover, GenAI's impact on \textit{equity and accessibility} has been questioned, with visually impaired users facing difficulties verifying content accuracy without additional support \citep{glazko2023autoethnographic}.

\textit{Ownership and intellectual property} issues can also challenge the applications of GenAI for educational purposes, especially in terms of generating images. The blurring of authorship lines can lead to disputes, particularly when GenAI is used to create deepfakes or plagiarised content, undermining the integrity of educational materials \citep{chiu2023impact}. While much of the existing research has focused on educators' concerns, little attention has been given to the ethical concerns that may be raised by students, who are key stakeholders in the learning process. Ensuring their perspectives are considered is crucial, particularly in the case of data comics, where ethical implications may directly impact their understanding and engagement \citep{ALFREDO2024100215}.

\subsection{Contributions and Research Questions}
This study aims to address the gaps identified above by applying principles for designing data comics with GenAI assistance and empirically evaluating their effectiveness in communicating data insights to students. Situated at the intersection of human behaviour and emerging technologies, this work builds on prior research exploring the potential of data storytelling to facilitate information retrieval and comprehension \citep{shao2024data, gomez2023personal,milesi2025piecing}, our contributions include a quantitative comparison of the \textbf{effectiveness} of data comics versus conventional visualisations in conveying data insights for \textbf{information retrieval} and \textbf{comprehension} tasks. These two tasks align with the foundational levels of Bloom's taxonomy as adapted for data visualisation \citep{Arneson2018bloom}, which have also been examined in prior comparative studies on conventional data visualisations \citep{shao2024data, milesi2025piecing}. We also investigate the \textbf{impact of students' visualisation literacy} levels on the effectiveness to which students complete these tasks with conventional visualisations and data comics, respectively. Additionally, we assess \textbf{students' perceptions} of the factors that may enhance or hinder insight communication, including perceived benefits and limitations. Lastly, we examine students' \textbf{ethical concerns} regarding the use of GenAI in assisting the creation of data comics.  
As a result, we propose the following research questions:

\begin{enumerate}[label=\textbf{RQ\arabic*.}, leftmargin=*, topsep=0.8em,]
    \item To what extent do data comics \textbf{effectively} communicate data insights to students compared to conventional visualisations? 
    \item To what extent do the effectiveness of data comics, as compared to conventional visualisation, vary in communicating data insights for \textbf{information retrieval} and \textbf{comprehension tasks}? 
    \item To what extent do the students' \textbf{visualisation literacy} influence the \textbf{effectiveness} of data comics compared to conventional visualisations?
    \item What are the students' \textbf{perceptions on potential benefits and limitations} of data comics in communicating data insights compared to conventional visualisations?
    \item What are students' perceived \textbf{ethical implications} of employing GenAI in supporting the creation of data comics?
\end{enumerate}

\section{Methods}

\subsection{Dataset and Materials} \label{canva}
For this study, we selected the globally relevant topics of climate change and global warming, which have garnered increasing concern worldwide. Some data in this field have proven difficult for students to interpret, especially those with limited visualisation skills \citep{sheppard2005landscape, gezer2022examination}. We included four conventional visualisations in our study: one based on a dataset from \textit{Kaggle} and the others derived from a pertinent comparative study \citep{shao2024data}.

To generate the four data comics, we drew on the established data comic design patterns proposed by \citet{bach2018design}, which structure data-driven stories along two key dimensions: \textit{content relation} (how individual insights relate to one another) and \textit{layout} (how panels are spatially arranged on the page). Guided by these patterns, we followed a two-stage process -- separating human-led narrative design from the human--AI co-creation of visual elements -- carried out by three researchers: one with expertise in data science and visual analytics (RA) and two with backgrounds in visualisation and dashboard design (RB \& RC). The full process, with worked examples and illustrative panels, is detailed in Appendix \ref{sec:datacomicprocess}.
 
In \textbf{Stage 1 (human-led narrative design)}, RA acted as the primary content creator, extracting key insights from the data and classifying each into one or more of the six content relation categories defined by \citet{bach2018design} (Narrative, Temporal, Faceting, Visual Encoding, Granular, and Spatial; see Appendix \ref{appendix:a6_content}). Based on these categories, the insights were mapped onto panel layouts (e.g., Parallel, Annotated) selected from the same set of design patterns, so that the sequencing and spatial arrangement of panels guided the reader's attention in a principled manner. RB and RC acted as reviewers, validating the chosen insights and contributing to decisions about narrative sequencing and layout structures.
 
In \textbf{Stage 2 (human--AI co-creation of comic strips)}, RA generated and refined image elements using \textit{DALL-E~3} (OpenAI), accessed via its integration with ChatGPT-4 and following the prompting guidelines reported in Appendix \ref{image creation}. The curated images were then integrated into the pre-defined panel layouts to construct the comic strips, with RB and RC providing input on visual consistency and narrative clarity across panels. Finally, storytelling elements such as titles, annotations, and highlights were layered onto the strips, and the comics were assembled using \textit{Canva}.

The final set of four data comics, created through the aforementioned crafting process, is presented in Appendix \ref{appendix:a3_comparative}.

\subsection{Study Design and Procedure\label{Study design}}
We designed a controlled study in which we manipulated a single variable: visualisation type -- the conventional visualisation condition (CV), versus the data comics condition (DC). A within-subjects study design was implemented, where participating students were presented with both conventional visualisations and data comics. To counterbalance its order effects, we randomised the order of visualisation presentation. Therefore, each participant served as their own control, ensuring the statistical power and sensitivity in investigating effects \citep{cohen1988statistical}.
The study was established in the form of a survey using Qualtrics, an online platform for designing and conducting research experiments. Ethics approval was obtained from the Human Research Ethics Committee at the university where the study was conducted (Project ID: 37307) and consent from participants were acquired. The study was structured to be accomplished within approximately 35 minutes. The study consisted of five parts (see details in Appendix \ref{appendix:a1_demographic} to \ref{appendix:a5_ethical}):
%https://docs.google.com/document/d/1DIUP9UCiLdJNsKOC3hrZZfFJyqcBtzk9/edit 

\begin{enumerate}
    \item Part 1: Demographic questions;
    \item Part 2: Basic visualisation literacy questions (relevant to RQ3) - the Mini-Visualisation Literacy Assessment Test (Mini-VLAT) developed by \citet{pandey2023mini} was employed to assess students' comprehension and interpretation of various visualisation types, including those used in the study (e.g., bar charts and stacked area charts);
    \item Part 3: Comparative visualisations questions (relevant to RQ1 and RQ2) - students were presented with four sets of questions, each containing two information retrieval questions and two comprehension questions;
    \item Part 4: Open-ended perception questions (relevant to RQ4) - students' perceptions and attitudes regarding using data comics for insight communication were gathered through open-ended questions;
    \item Part 5: Ethical implication questions (relevant to RQ5) - open-ended questions were provided, allowing students to elaborate on their ethical concerns about using GenAI in the creation of data comics.
\end{enumerate}

Further details of the last three sections are provided next. 

\subsubsection{Comparison Study (Study Part 3)} \label{comparison_study}

\textbf{Data Comics vs. Conventional Visualisations}. 
This subsection outlines Part 3 of our study, where we created four pairs of visualisations - four conventional visualisations (CV1-4) and four data comics (DC1-4). Each pair comprised a conventional visualisation and a data comic, both presenting the same data insights under different conditions to maintain comparison consistency. 

\begin{figure}[!htbp]
    \centering
    \includegraphics[width=0.95\linewidth]{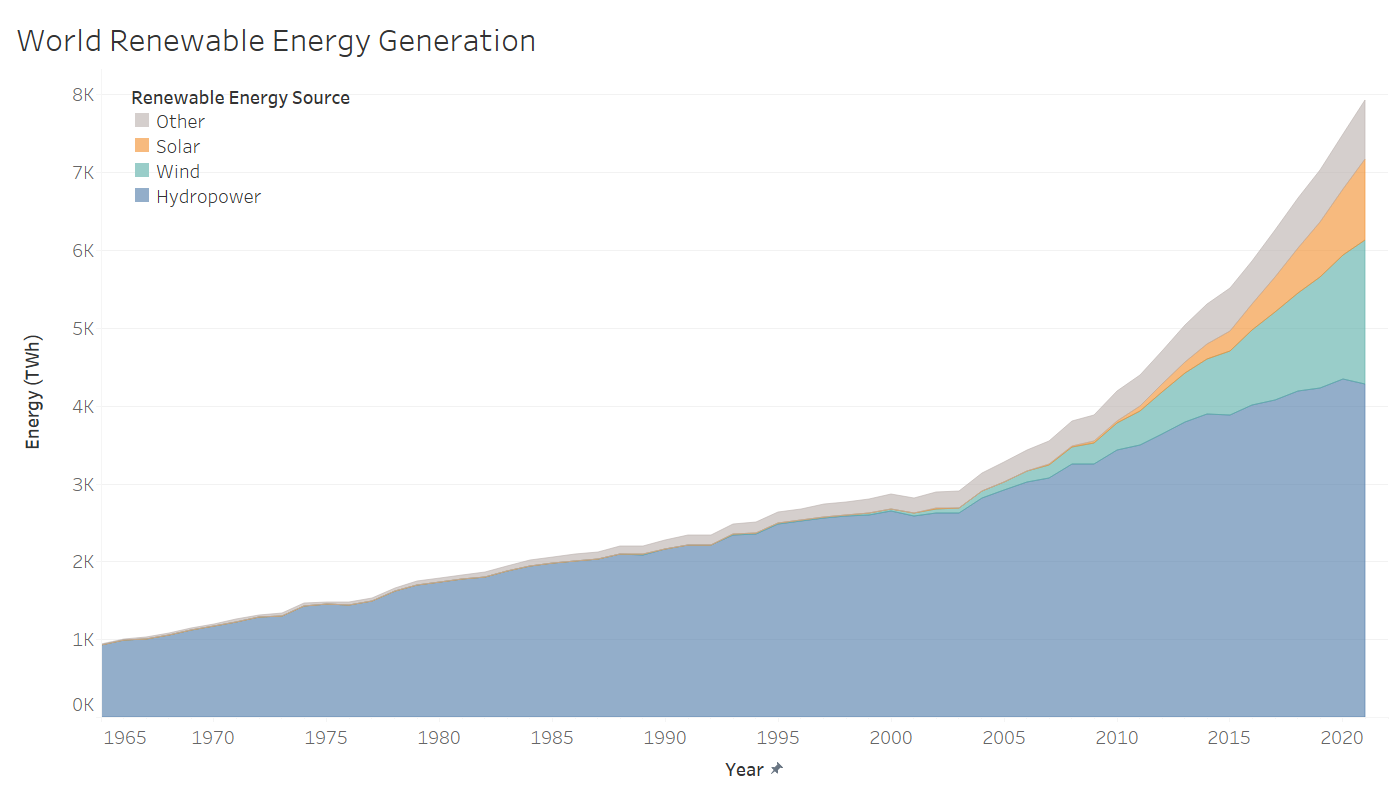}
    \caption{Conventional visualisation - CV2 in comparative pair 2}
    \label{fig:CV2}
\end{figure}

\begin{figure}[!htbp]
    \centering
    \includegraphics[width=0.79\linewidth]{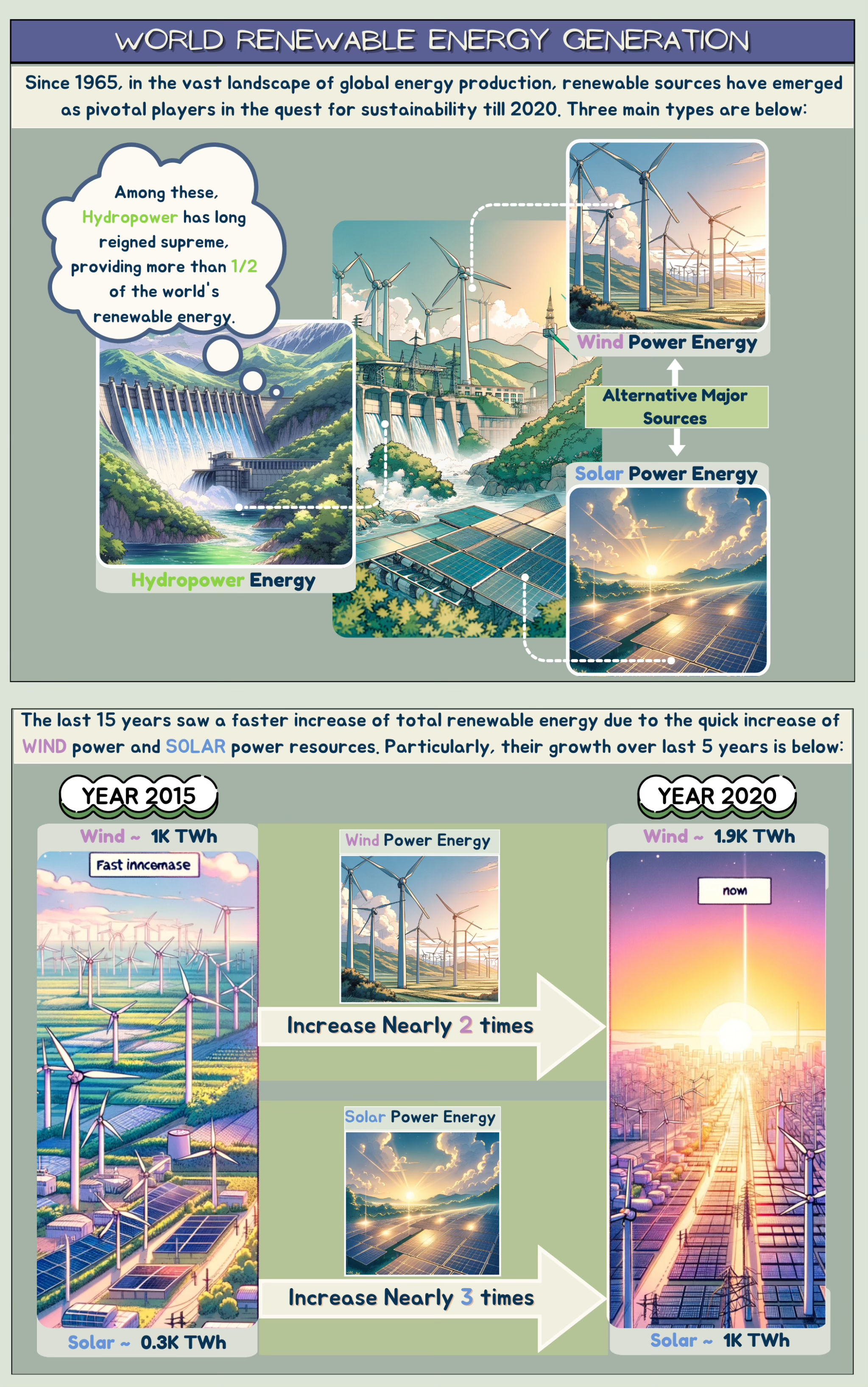}
    \caption{Data comic - DC2 in comparative pair 2}
    \label{fig:DC2}
\end{figure}

Figures \ref{fig:CV2} and \ref{fig:DC2} show an example of one of our comparative pairs, specifically pair 2, which includes CV2 and DC2, respectively. To mitigate order effects and learning biases in our study, we used a counterbalanced design where each participant randomly viewed two items from each condition in a random order, unaware of which were conventional or comics. For example, one participant might view visualisations CV1 and CV2 along with comics DC3 and DC4, while another might encounter CV3 and CV4 paired with comics DC1 and DC2 and so on.

\textbf{Types of questions}. 
Following the approach by \citet{shao2024data} and \citet{milesi2025piecing} for scrutinising the effectiveness of data storytelling approaches, we structured the questions for our comparative visualisations using the first two levels of Bloom's taxonomy tailored to data visualisation tasks \citep{Arneson2018bloom,byrd2019using}: ``knowledge'' and ``comprehension''. These levels were chosen to align with the goals of our RQ2. Participants were presented with four questions per visualisation — two assessing their ``knowledge'' and two assessing their ``comprehension''. For consistency, the same set of questions was used for each paired visualisation (e.g., CV1 and DC1). The design of each question category is detailed below:

Level 1 - \textbf{Knowledge} questions in Bloom's taxonomy served to assess whether users can identify and select relevant data points, categories, or trends from a dataset. In our study, participants were required to answer two information retrieval questions for each visualisation (see Table \ref{question type sample}, row 1). 

Level 2 - \textbf{Comprehension} questions tested the participants' abilities to understand, interpret, and derive insights from visualisations. Participants were asked to respond to two comprehension questions per visualisation. These questions were split into two sub-types: i) single insight - required selecting the correct answer about a single insight provided by the visualisation (see Table \ref{question type sample}, row 2), and ii) multiple insights - involved identifying an incorrect statement among four, requiring a broader understanding of multiple insights \citep{Arneson2018bloom,byrd2019using} (see Table \ref{question type sample}, row 3). 

\begin{table}[h]
\caption{Examples of information retrieval and comprehension questions asked for both conventional visualisations and data comics conditions}
\resizebox{\columnwidth}{!}{%
\begin{tabular}{l|l|l}
\toprule
\textbf{Bloom Taxonomy}                 & \textbf{Question Type (each visualisation)} & \textbf{Examples}                                                                                                                                                                                                                                                                                         \\ \midrule
Level 1: Knowledge                      & Information Retrieval (x2)                  & \begin{tabular}[c]{@{}l@{}}Since which year did the total renewable energy see a faster increase?\\ 1. 1965\\ 2. 2000\\ 3. 2005\\ 4. 2020\end{tabular}                                                                                                                                                    \\ \midrule
\multirow{2}{*}{Level 2: Comprehension} & Comprehension (single insight x1)              & \begin{tabular}[c]{@{}l@{}}
What may be the percentage of Agriculture lands by 2000?\\ 1. 5\%\\ 2. 9\%\\ 3. 22\%\\ 4. 47\%
\end{tabular} \\ \cmidrule{2-3} 
                                        & Comprehension (multiple insights x1)           & \begin{tabular}[c]{@{}l@{}}Please choose the sentence that is incorrect: \\ 1. Oil emissions equal to emissions from coal in 2004\\ 2. Oil and gas were massively used from the early 21st century\\ 3. Coal has been used since the 1800s\\ 4. Oil emissions increased rapidly before 1980s\end{tabular} \\ \bottomrule
\end{tabular}%
}

\label{question type sample}
\end{table}

\subsubsection{Open-ended Perception Questions (Study Part 4)}
In this part of the study, participants were invited to share their perspectives on the effectiveness of data comics in conveying data insights compared to conventional visualisations by responding to four open-ended questions. We gathered participants’ views on the features (if any) of data comics that may enhance communication and engagement, their examples of instances where data comics may have outperformed conventional visualisations (if this was perceived), and their perceived limitations or concerns, addressing RQ4 (see questions in Appendix \ref{appendix:a4_perception}). %by asking: i) What specific features of data comics contribute to their effectiveness in communicating data insights compared to conventional visualisations? Elaborate on how the feature enhances the effectiveness; ii) What specific features of data comics impact their effectiveness in engaging audiences compared to conventional visualisations? Elaborate on how the feature enhances the effectiveness; and iii) Conversely, are there any limitations or drawbacks of data comics that you perceive as hindering their effectiveness in communicating data insights compared to conventional visualisations? If so, what are they? 
Their responses were then analysed to shed light on investigating the advantages and disadvantages of data comics, addressing our RQ4.

\subsubsection{Ethical Implication Questions (Study Part 5)}
In this final part of the study, we revealed to students that all data comics presented in the previous questions were produced to some extent with the assistance of GenAI. We provided three yes/no questions to allow students to express their ethical concerns regarding the use of GenAI in the creation of visual content, specifically focusing on commonly-held concerns regarding \textit{misinformation} \citep{yan2024vizchat}, \textit{bias} \citep{farrelly2023generative}, and \textit{ownership} \citep{chiu2023impact} as discussed in Section \ref{ethical_concerns}, addressing RQ5 (see in Appendix \ref{appendix:a5_ethical}). Two open-ended questions were provided to elaborate on these three ethical concerns and express additional concerns beyond those mentioned above. For example, one question was: \textit{``Do you believe there are concerns about potential misinformation associated with the use of GenAI-powered data comics?'' If yes, please elaborate on your concerns}. 

\subsection{Participants}

%Participant recruitment was conducted through two channels: i) Prolific, an online platform supporting academic research that can be used to recruit students across the globe; and ii) personal connections, specifically participants studying at the university where the researchers were based in. An a priori power analysis using G*Power 3.1 \citep{faul2009statistical} indicated that a sample size of 47 participants was required to achieve sufficient statistical power ($\alpha = 0.05, \text{power} = 0.95$) with an effect size of 0.5, based on \citep{shao2024data,milesi2024qualitative}. We ultimately recruited 60 student participants. Eligibility criteria required participants to have advanced English proficiency and to complete the survey on a laptop or desktop to ensure consistent visualisation displays. 
%Prior to the main study, we conducted two pilot studies with two participants to refine the design; their responses were not included in the final analysis. Participants were permitted to skip at most three questions; skipping more than three, or failing to complete the survey, resulted in their responses being excluded. 
%Before rolling out the actual study, we conducted two pilot studies. We involved two participants to fine-tune the study and their responses were not incorporated into the main study. Participants were allowed to skip at most three questions, and skipping more than three or not completing the whole survey would result in an invalid response and exclusion. 
Participant recruitment was conducted through two channels: Prolific, an online platform for academic research, and the researchers' university. An a priori power analysis using G*Power 3.1 \citep{faul2009statistical} indicated that 47 participants were required to achieve sufficient power ($\alpha = 0.05$, power = 0.95, effect size = 0.5), based on \citep{shao2024data,milesi2024qualitative}. We recruited 60 student participants. Eligible participants were required to have advanced English proficiency and complete the survey on a laptop or desktop to ensure consistent visual display. Before the main study, two pilot participants helped refine the survey; their responses were excluded from analysis. Responses were considered invalid if participants skipped more than three questions or failed to complete the survey. The participants' demographics were reported in Appendix \ref{appendix:a7_demographics}.

\subsection{Analysis} 
\subsubsection{Metrics\label{Metrics}}
We defined the following metrics to address our research questions and both of them were calculated for both conventional visualisations and data comics:
\begin{itemize}
    \item \textbf{Correct Rate}:
    In the comparison study detailed in Section \ref{comparison_study}, each question regarding individual pairs of visualisations carried a raw score of 1 for each correct answer. For calculating the Correct Rate (CR) based on each participant, we collected associated indices: i) the number of total questions (TQ), which was 16 for each condition (CV and DC) and ii) the total number of correct answers (CA). The computing formula is given below:
    
\begin{equation}
    CR = CA/TQ
\end{equation}

To provide a more accurate reflection of the individual's actual understanding, minimising the impact of guesses on their performance, we implemented the ``correction for guessing'' formula proposed by \citet{frary1988formula}. This formula is employed to adjust scores in assessments, particularly suitable for multiple-choice questions, reducing the influence of guessing \citep{frary1988formula}:

\begin{equation}
    CS = Right - Wrong/(k-1)
\end{equation}

In this formula, ``CS'' represents the correct rate of a participant's accuracy (corrected for guessing). ``Right'' refers to the total number of correct answers, while ``Wrong'' reflects the total number of incorrect answers. The variable ``k'' indicates the number of choices per question, set at four in our study. By applying these formulas in our case, the adjusted CS was used to account for the correct answer (i.e., CA = CS), and hence, the numerical value for Correct Rate (CR) ranges from 0 to 1.

\end{itemize}

\begin{itemize}
    \item \textbf{Visualisation Literacy Correct Rate}: Similar to the calculation of Correct Rate detailed in Section \ref{Metrics}, each question in the Mini-VLAT (see in Section \ref{Study design}) also carried 1 raw score for each correct answer. For calculating the Visualisation Literacy Correct Rate (VLCR) based on each participant, we collected associated indices: i) the number of total questions in Mini-VLAT(VLTQ), which was 12 and ii) the total number of correct answers in Mini-VLAT (VLCA). The computing formula is given below:
    
\begin{equation}
    VLCR = VLCA/VLTQ
\end{equation}
We also applied the ``correction for guessing'' formula \citep{frary1988formula} to adjust for the probability of correct answers that may have been achieved through guessing. The value range for this metric is 0 to 1.
\end{itemize}

\subsubsection{Analysis for each RQ}  \label{analysis}
To address \textbf{RQ1}, we utilised R to compute the median and interquartile range (IQR) of the correct rate metric, visualising the results with a box plot. Specifically, we conducted the Shapiro-Wilk W-test to evaluate the normality of the data collected \citep{royston1992approximating}. Our analysis revealed that the correct rates collected in our study did not follow a normal distribution ($p < 0.05$), and the visualisation literacy correct rates were likely not normally distributed (p $<$ 0.1). Therefore, we applied the Wilcoxon signed-rank test, tailored for paired data and accommodating individual differences by regarding each participant as their own control. %This analysis was used for determining whether the correct rate metric in data comics significantly differed from that observed in conventional visualisations for the same participants. 
In our analysis, we adopted a significance threshold of p $<$ 0.05 to assess statistical significance \citep{nachar2008mann}. We used the $r=(Z/\sqrt{N})$ as the effect size measure, chosen for its appropriateness in comparing non-parametric paired data. 

To address \textbf{RQ2}, we categorised the collected data based on question type, distinguishing between information retrieval and comprehension tasks. For comprehension tasks, we further identified subtypes, containing single insights versus multiple insights. For each subtype, we computed the median and interquartile range (IQR), visually presenting these statistics through box plots. Furthermore, we employed the same statistical tests used for RQ1 to evaluate differences in effectiveness between data comics and conventional visualisations across the two task levels. To prevent the issue of statistical significance occurring by chance due to multiple comparisons, we applied Bonferroni correction to adjust the p-values by multiplying the original values by 4 (\textit{n} comparisons) \citep{jafari2019and}.

To address \textbf{RQ3}, we tested whether visualisation literacy (VL) predicts task accuracy and whether this relationship differs between data comics (DC) and conventional visualisations (CV) conditions. We conducted a linear regression with \textit{CorrectRate} as the dependent variable, and VL correct rate, Condition and their interaction (\textit{VL correct rate × Condition[DC]}) as predictors. This interaction tested whether the effect of VL on accuracy varied across conditions.

%to examine VL performance (Mini-VLAT correct rate), investigating the correlation between the participants' VL and effectiveness in different conditions, measured as the \textit{CorrectRate}. Furthermore, we included an interaction term named (\textit{minivalt × Condition[DC]}) to evaluate whether the effect of Mini-VLAT on correct rate impacted by each condition and vise versa. 
In all models, $\beta_0$ referred to the intercept, with the CV condition serving as the baseline, while Condition[DC] indicated the DC condition.

\begin{equation}
    \begin{split}
        CorrectRate &= \beta_0 + \beta_1 \times minivlat + \beta_2 \times Condition[DC] \\
        &\quad + \beta_3 \times (minivlat \times Condition[DC])
    \end{split}
\end{equation}

To address \textbf{RQ4}, two independent coders carried out thematic analyses \citep{braun2006using} on participants' perceptions concerning the advantages and disadvantages of using data comics for conveying data insights versus conventional visualisations, coding each of the 60 valid responses\footnote{While the participants were asked to answer four questions, the last question was optional. Thus, participants' answers to the four (or three in some cases) questions were concatenated as a single response per participant.}. During the analyses, emerging themes were validated through an iterative process of constant comparison with the raw data and refinement. Reflexive discussions among the research team helped to check potential biases and ensure that interpretations remained grounded in participants' accounts. Final themes were assessed collaboratively for coherence and distinctiveness, with illustrative quotations used to demonstrate their credibility. The consistency of the thematic analysis was validated by assessing inter-rater reliability through Cohen's Kappa \citep{cohen1960coefficient}. In these analyses, the two coders achieved Cohen's kappa values of 0.76 and 0.81 for benefits and limitations respectively, indicating ``substantial'' to ``almost perfect'' agreement \citep{lazar2017analyzing}\footnote{\citet{lazar2017analyzing} described kappa values between 0.61 and 0.80 as ``substantial'' agreement, and values between 0.81 and 1.00 as ``almost perfect'' agreement.}. Disagreements were resolved through case-by-case discussion between the two coders to enhance consistency.

 %The assumptions of homoscedasticity, normality, independence of residuals, linearity, and multicollinearity were confirmed for all regressions.

%\begin{equation}    
    %CorrectRate = \beta_0 + \beta_1 \times data\_literacy + \beta_2 \times Condition[DC] + \beta_3 \times (data\_literacy \times Condition[DC])
%\end{equation}
%\begin{equation}
    %CorrectRate = \beta_0 + \beta_1 \times vis\_literacy + \beta_2 \times Condition[DC] + \beta_3 \times (vis\_literacy \times Condition[DC])
%\end{equation}

To address \textbf{RQ5}, we quantified the frequency of participants' concerns regarding leveraging GenAI in the production of data comics, with the primary focus on misinformation, bias, and ownership issues. For the two open-ended questions, we combined their answers into a single entry to capture each participant's perspective holistically; thus, the participant is the unit of analysis. We then applied thematic analysis \citep{braun2006using}, following an inductive coding approach to identify recurrent patterns in participants' concerns. Codes were iteratively refined into broader themes that captured common perspectives and issues raised across participants. Responses from the sixty participants were coded independently by two researchers using broader themes, and inter-rater reliability was evaluated to ensure consistency and reliability in the analysis of this segment \citep{nowell2017thematic}. The two coders achieved a Cohen's $\kappa$ of 0.83, indicating a very good level of agreement in evaluating participants' ethical concerns.

\section{Results}
\subsection{Preliminary Exploration}
Before diving into the research questions, a preliminary analysis of participants' responses in Study Part 3 (Section \ref{comparison_study}) was conducted. Our exploration revealed that participants' overall accuracy was moderate (median: 0.500, IQR: 0.250). Our analysis showed that participants performed better on \textbf{information retrieval questions} (median = 0.667, IQR = 0.333) than on \textbf{comprehension questions} (median = 0.333, IQR = 0.333). This difference was significant, with a large effect size (Mann--Whitney $U = 2982.5$, $Z = 6.332$, $p < 0.001$, $r = 0.578$). Within comprehension items, performance was comparable between \textbf{single-insight} (median = 0.333, IQR = 0.667) and \textbf{multiple-insight} questions (median = 0.333, IQR = 0.334). The difference was small and only marginally significant ($U = 1445.0$, $Z = -1.971$, $p = 0.05$, $r = 0.18$). However, performance on information retrieval items remained significantly higher than on both single- and multiple-insight comprehension questions ($p < 0.001$). These preliminary findings suggest that comprehension questions were consistently more challenging than retrieval questions, as intended in the study design.

\subsection{Effectiveness of Data Comics (RQ1)}
\begin{figure}[!htbp]
    \centering
    \includegraphics[width=0.7\linewidth]{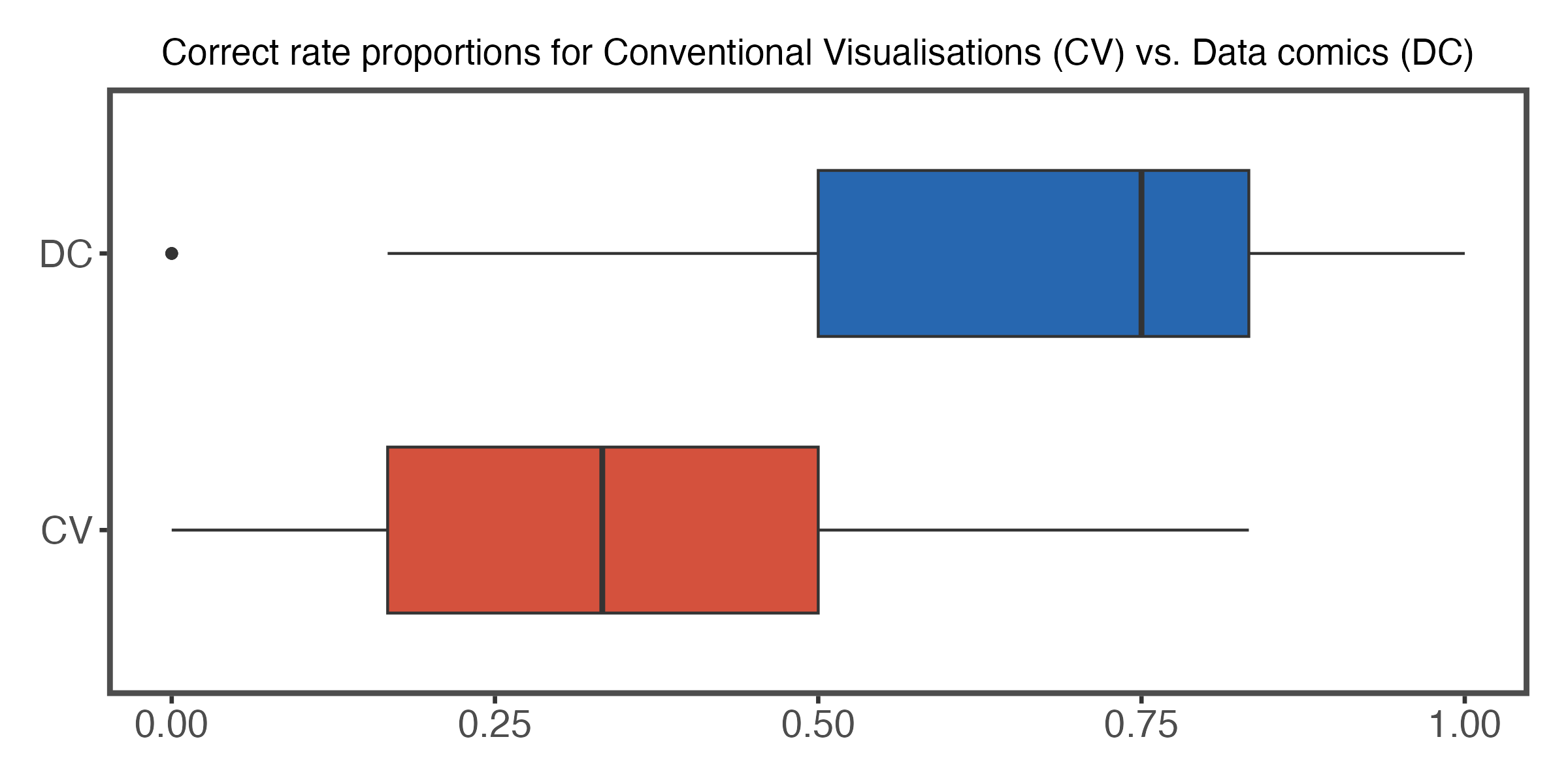}
    \caption{Correct rate of data comics (DC) versus conventional visualisations (CV). The difference was significant (p $<$ 0.001).}
    \label{fig:Results for RQ1}
\end{figure}

To address \textbf{RQ1}, we compared the effectiveness of data comics (DC) and conventional visualisations (CV) in communicating data insights, measured by participants' \textit{Correct Rate} (correct answers). participants achieved a significantly higher \textit{CorrectRate} with DC (median = 0.750, IQR = 0.333) than with CV (median = 0.333, IQR = 0.333), indicating a large effect ($U = 1705.5$, $Z = 5.143$, $p < 0.001$, $r = 0.664$). These differences are illustrated in Figure \ref{fig:Results for RQ1}. 
These findings suggest that data comics enhance accuracy, highlighting the potential of data comics to improve the communication of insights. Further analyses are required to examine differences across specific question types, which we present in the next section.

\subsection{Effectiveness in Information Retrieval and Comprehension Tasks (RQ2)}

To address RQ2, we examined the impact of data comics across two task levels: information retrieval and comprehension of insights, further dividing comprehension into single-insight and multiple-insight tasks.

\begin{figure}[!htbp]
    \centering
    \includegraphics[width=0.7\linewidth]{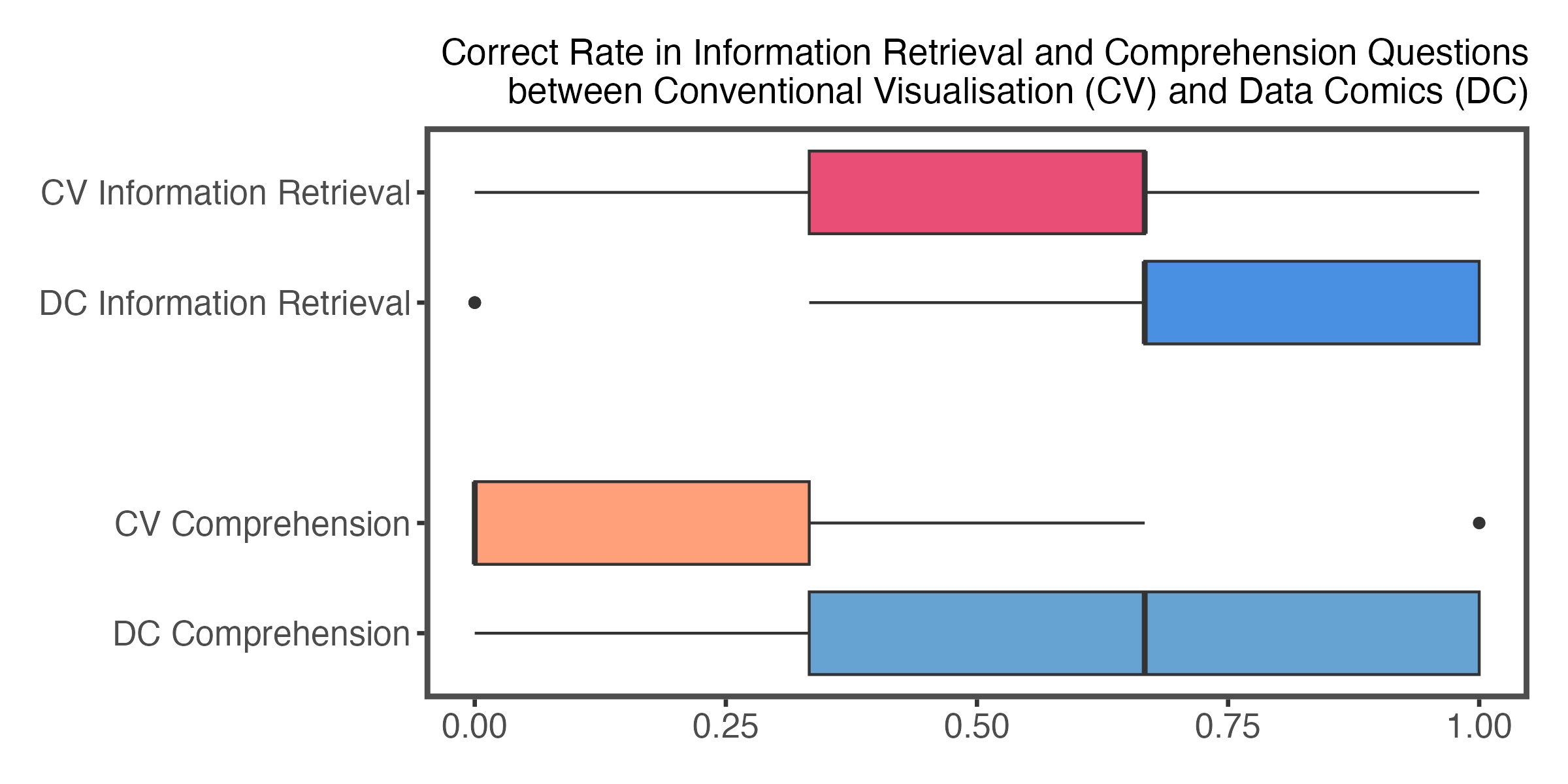}
    \caption{Correct rate in Information Retrieval and Comprehension tasks between conventional visualisations (CV) and data comics (DC). Participants presented with DC achieved higher correct rates in Information Retrieval and Comprehension tasks compared to those with CV, with the difference reaching statistical significance (p $<$ 0.05).}
    \label{fig:Results_RQ2_K_C}
\end{figure}

For information retrieval tasks, participants achieved a significantly higher correct rate in the DC condition (median = 0.667, IQR = 0.333) compared to conventional visualisations (CV) (median = 0.667, IQR = 0.334), with a moderate effect size ($U = 1601.5$, $Z = 2.642$, $p = 0.033$, $r = 0.341$), corresponding to an average increase of 0.156. 
For comprehension tasks, participants also performed significantly better with DC (median = 0.667, IQR = 0.667) than with CV (median = 0.000, IQR = 0.333), showing a large effect size ($U = 1727.5$, $Z = 5.383$, $p < 0.001$, $r = 0.695$). That is, data comics increased the correct rate by a mean of 0.417, suggesting greater benefits for comprehension relative to retrieval tasks. These differences can be seen in Figure~\ref{fig:Results_RQ2_K_C}. 

\begin{figure}[!htbp]
    \centering
    \includegraphics[width=0.7\linewidth]{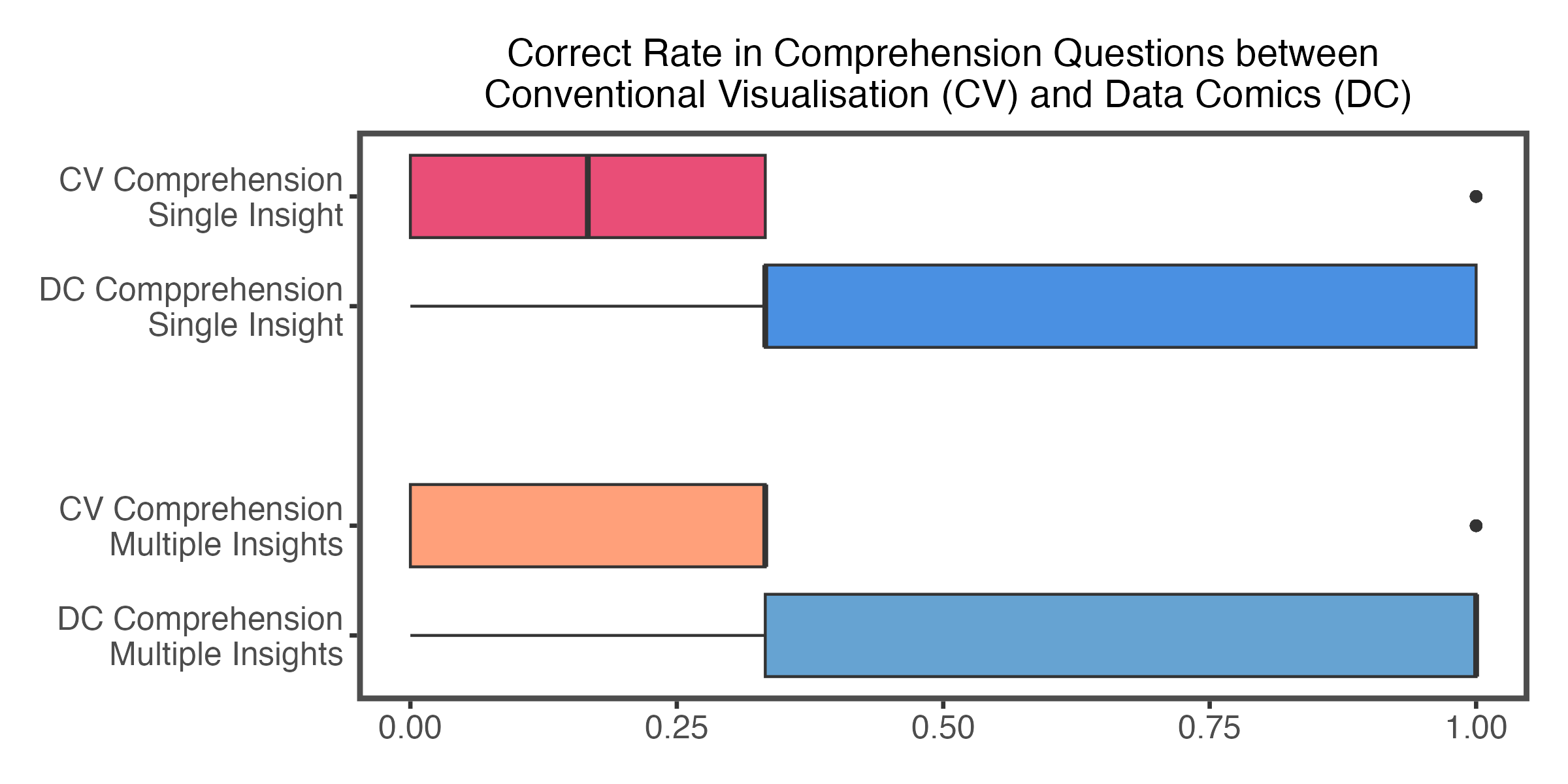}
    \caption{Correct rate in Comprehension tasks with single and multiple insights between conventional visualisations (CV) and data comics (DC). Participants presented with DC achieved higher correct rates in comprehension tasks compared to those with CV regardless of the number of insights, with the difference reaching statistical significance (p $<$ 0.001).}
    \label{fig:Results_RQ2_Insights}
\end{figure}

When separating comprehension into subtypes, DC also outperformed CV. For \textbf{single-insight} questions, DC yielded a higher correct rate (median = 0.333, IQR = 0.667) than CV (median = 0.167, IQR = 0.333), with a large effect ($U = 1676.0$, $Z = 4.030$, $p < 0.001$, $r = 0.520$), an average increase of 0.322 (Figure~\ref{fig:Results_RQ2_Insights}). For \textbf{multiple-insight} questions, DC showed an even larger improvement (median = 1.000, IQR = 0.667) compared to CV (median = 0.333, IQR = 0.333), with a large effect ($U = 1690.0$, $Z = 4.564$, $p < 0.001$, $r = 0.589$), corresponding to an average increase of 0.406 (see Figure~\ref{fig:Results_RQ2_Insights}).  
 
In summary, across both information retrieval and comprehension tasks, data comics consistently outperformed conventional visualisations, with the largest gains observed in comprehension, particularly for multiple-insight questions, highlighting their potential to support deeper data understanding.

%In sum, we found consistent evidence in data comics' favour for both information retrieval and comprehension tasks compared to conventional visualisations, particularly manifest in comprehension tasks, thereby emphasising its potential critical role in enabling insightful data understanding.

\subsection{Visualisation Literacy and Effectiveness (RQ3)}

To address \textbf{RQ3}, we examined the relationship between visualisation literacy (VL) and effectiveness using a linear regression model (see Equation~(4) in Section~\ref{analysis}). The model explained a significant proportion of variance ($R^2 = 0.37$, $F(3,116) = 22.82$, $p < 0.001$). The intercept was significant ($\beta_0 = 0.172$, $t(116) = 2.359$, $p = 0.020$), indicating a baseline correct rate of 17.2\% when VL = 0 and the condition was conventional visualisation (CV). VL was a significant positive predictor of effectiveness ($\beta_1 = 0.371$, $t(116) = 2.909$, $p = 0.004$), suggesting that each 1\% increase in VL corresponded to a 0.371\% increase in task accuracy.  

 We also found a significant main effect of condition ($\beta_2 = 0.281$, $t(116) = 2.717$, $p = 0.008$), indicating that participants working with data comics (DC) achieved on average 28.1\% higher accuracy compared to CV. However, the interaction term (\textit{VL $\times$ Condition[DC]}; $\beta_3 = 0.059$, $t(116) = 0.327$, $p = 0.745$) was not significant, indicating that the positive association between VL and effectiveness was consistent across both conditions.  

 %This indicates that we found no evidence of an interaction effect, meaning the relationship between visualisation literacy and effectiveness appeared consistent across both conventional visualisations and data comics.
%In addition, we observed a significantly positive correlation between participants' effectiveness and the conditions, i.e., whether they work on conventional visualisation or data comics, ($\beta_2=0.281, t(116)=2.717, p=0.008$). It suggested that participants, when working on questions with insights delivered in the form of data comics, achieved 28.1\% higher in tasks performance compared to conventional visualisations. The interaction term (\textit{minivlat × Condition[DC]}; $\beta_3=0.059, t(116)=0.327, p=0.745$), however, was not statistically significant, indicating that the increase of effectiveness due to the increase of individuals' visualisation literacy was not affected by the visualisation condition they were in, and vice versa. 

%Together with the results from RQ1 and RQ2, this finding highlighted that data comics had a positive impact on the effectiveness of data insights communication, allowing participants to perform better even with varying levels of visualisation literacy, although ultimately visualisation literacy is critical to make the most of both conventional data visualisations and data comics.

Taken together, these results indicate that while data comics positively impact the communication of data insights (RQ1), particularly for higher-level comprehension tasks (RQ2), visualisation literacy remains a critical factor (RQ3). Even in the data comics condition, students with higher visualisation literacy outperform those with lower literacy, much as they do when using conventional visualisations.

\subsection{Perceived Benefits and Limitations of Data Comics (RQ4)}

\subsubsection{Perceived Benefits} 
All participants agreed that data comics offered notable benefits (N=60). Ninety percent (N=54) perceived them as \textit{more accessible} for broader audiences and more efficient in conveying information compared to conventional visualisations. Elaborations on perceived benefits were distilled in the following three themes. 

\textbf{\textit{Theme 1 - Accessible and efficient insight communication.}} The first theme referred to the perceived augmented accessibility and efficiency of data comics in delivering data insights (N=54). For some participants, data comics can make data relatable as they ``\textit{contextualise data}'' (P15, P23) through ``\textit{weaving the story}'' (P50) behind the data, offering an ``\textit{easier way to associate the data to the corresponding topic}'' (P46). The assembling of pictorial and textual elements was perceived as an organised reading sequence (``\textit{works like a comic book to convey information step-by-step}'' - P4), leading participants to ``\textit{easily follow the flow of data storytelling}'' (P8). With the assistance of visual elements (e.g., bold fonts or explanatory annotations), some participants said that they found data comics to be ``\textit{easier to follow and understand}'' (P4, P45, P57), thus audiences were ``\textit{quicker to grasp the key point}'' (P6, P7). A few participants (P5, P21) perceived it easier to differentiate data relationships (``\textit{pictures and textual explanations allowed me to know the relationship more directly than traditional data visualisations}'' - P14). Notably, some participants (P4, P39, P53) also described data comics to be ``\textit{easier for people who have difficulty reading graphs}''. These comments suggest that data comics, in contrast to conventional visualisation, provide more accessible opportunities in conveying intended data insights efficiently.

\textbf{\textit{Theme 2 - Engaging presentation.}} Another theme identified referred to participants' perceiving data comics as a more attractive format to engage the audience into the story (N=44). The majority of participants expressed positive views about the use of colourful images and visual cues (N=36), which were depicted to be fairly appealing and attractive (``\textit{Comics are sort of fun and captivating, which keeps audience engaged} - P40''). These graphics ``\textit{make the data more emotive}'' (P45, P18) and hence, further ``\textit{awakening unmotivated readers}'' (P52). Moreover, the manner in which information is laid out was perceived as contributing to foster ``\textit{more entertaining reading}'' (P20), which was echoed by the perspective of one student arguing that ``\textit{The visual diversity of data comics is better at capturing and holding the viewers' attention than cookie-cutter statistical graphics}'' (P14). Interestingly, one student noted that data comics surpassed at ``\textit{associating with the traditional comic strip format to encourage the audience engage with them because they are familiar [with comics]}'' - P31. These comments suggest that data comics can enhance the aesthetic appeal of information presentation to create a more engaging experience for the audience. 

\textbf{\textit{Theme 3 - Perceived ease of information retention.}} Some participants (N=20) reported finding it easier to recall key insights in data comics conditions. The cartoonish style was perceived as more interesting to aid participants remembering key data points (P8, P10). Most participants pointed out that the imagery elements can promote the retention of detailed information (``\textit{visual metaphors and symbols which connects data points in a coherent story helping to maintain the reader's interest and improving memory retention}'' - P37; `` \textit{Images and pictures make it stand out more and give more retention}'' - P53). In addition, one student highlighted the value of the narrative style in boosting information recall, ``\textit{increasing the likelihood that readers will see the data presentation through to the end}'' - P60. These perspectives suggest the potential of data comics in enhancing memory retention through visual formats that associate cohesive narrative, making information more memorable. Nonetheless, we did not measure this construct.

\subsubsection{Perceived Limitations}
Two-thirds of participants (N=36) highlighted limitations in data comics, primarily citing information overload (N=18). Their responses were summarised in four themes.

\textbf{\textit{Theme 1 - More information = overwhelming.}} This theme referred to participants' perceived information overload in data comics to hinder efficient information location (N=18). Though the use of vividly coloured graphs captivated participants' interest in investigating the narrative, some found the layouts ``\textit{too busy}'' (P55), which may ``\textit{introduce confusion to the viewer}'' (P17), ``\textit{draw attention away}'' (P16), and ``\textit{affect locating key information}'' (P7), thus leading to more distractions than conventional visualisations. This sentiment was supported by P33, who stated: ``\textit{it's easy to add too much information, or lose track of the story}''. Additionally, a few participants also pointed out that data comics are more time-consuming in key insights locating as ``\textit{images took a lot of space}'' (P9), which ``\textit{slows down the reading pace}'' (P12). These comments indicate that data comics can sometimes increase readers' cognitive load with excessive detail, complicating the process of extracting key data insights. 

\textbf{\textit{Theme 2 - Less generalisable application.}} This theme suggested that data comics could fall short in depicting complex and diverse information, being inappropriate in some formal contexts (N=18). Occupied with visual elements and annotations, data comics suffer from ``\textit{limited space}'' (P3) to convey nuanced insights. Importantly, some participants (P5, P38, P57) pointed out the potential difficulty of comparing the multi-dimensional data relationships through data comics, a task that could be achieved by conventional visualisations (`` \textit{It may be difficult to compare large variations of data using data comics than it would be using conventional visualisations}'' -P57). Furthermore, P12 offered a critical viewpoint, identifying the format of data comics as ``\textit{informal or unsophisticated in professional or academic situations where the tone is serious}'', a sentiment echoed by another student (P47) who noted that ``\textit{conventional visualisations can get more of a global approach to the topic}''. These perspectives reveal that data comics may not be universally applicable, especially in settings requiring thorough data analysis. 
 
\textbf{\textit{Theme 3 - More design-centric and author-guided.} } Participants (N=15) perceived the information delivery of data comics to be highly dependent on how they were designed by the authors. Some participants (P14, P32, P34) perceived that producing comics was a labour-intensive task that necessitated relatively more time to design (``\textit{I reckon the distribution of data, images, text, and colours on a comic takes time to design}'' - P14). They noted that realising a suitable arrangement of pictorial and textual elements ``\textit{heavily relies on the design skills of the author}'' (P27). Poor design can adversely introduce ambiguity and misinterpretation of intended insights (P37, P46, P48). Moreover, several participants (P4, P18, P31, P56) also critiqued that the narrative approach in data comics could restrict participants' ability to independently interpret and creatively engage with data (``\textit{you are guiding the narrative directly and limiting the audience's ability to make their own opinions or extract their own insights}'' - P31). These comments suggest that, unlike conventional visualisations -- often perceived as more objective and open for independent exploration -- data comics are highly design-driven and thus more susceptible to the creator's choice of focus leaving little space for interpretation.
%These comments suggest that data comics, unlike conventional visualisations, place a higher premium on design expertise for information conveyance.

\textbf{\textit{Theme 4 - Partial information presentation and oversimplification.}} This theme suggested that data comics tend to present information selectively rather than providing a comprehensive view of the data (N=10). Participants (e.g., P23, P33, P37) pointed out that data comics might prioritise readability by simplifying complex information, presenting only a portion of the overall picture, which could result in the loss of precision. One student expressed concern that such simplification risks ``\textit{potentially omitting crucial details needed for deeper understanding}'' (P18). Similarly, others (P45, P49) pointed out that by ``\textit{not giving out the full picture of information to the viewers}'', data comics might ``\textit{lose a lot of data when only key figures are used}''. These perspectives underscore the importance of carefully balancing readability and the depth of insight when creating data comics, to ensure that the narrative remains accessible without sacrificing critical information.

\subsection{Ethical Concerns of Using GenAI}
A majority of participants (N=45) raised concerns about misinformation in using GenAI to support data comics creation, with two-thirds (N=40) highlighting bias and inequity issues. Additionally, many (N=37) viewed these aesthetics as a threat to intellectual property, and over one-third (N=23) perceived them as less reliable. Four themes captured these concerns.

\textbf{\textit{Theme 1 - Reinforcing misinformation.}} Participants were consistently concerned about the GenAI-generated graphics presented in data comics to inadvertently disseminate inaccurate information (N=45). Not limiting to the flaws of training data, such as ``\textit{being fed with incorrect information}'' (P22) or embedded ``\textit{algorithmic biases}'' (P23), ``\textit{generative models were concerned to be making mistakes}'' (P30), leading to the propagation of misinformation through data comics. Hallucination, a notable phenomenon of misinformation, was identified by a few participants as a potential source of images representing  ``\textit{nonfactual situation of the reality}'' (P7, P25). One student elaborated on this as follows: ``\textit{It's easy to tweak the input requests to get a false and faulty output out of them and yet the ultimate output image would look genuine}'' (P49). In addition, another student (P56) also questioned the overall quality of the images produced by GenAI, arguing ``\textit{I don't think the quality of AI-generated data comics would be high due to its low accuracy}''. These comments emphasised the scepticism surrounding the reliability and integrity of using GenAI applications in data comics.

\textbf{\textit{Theme 2 - Perpetuating biases and undermining fairness.}} This theme referred to the inevitable biases in output generated by GenAI, as observed by forty participants (N=40). AI models trained with datasets harbouring historical biases (P4, P22, P26), algorithmic biases (P23, P25), stereotypes (P3, P43), and skewed perspectives (P5, P40, P56), were likely to perpetuate portrayal or biased narratives and misrepresenting certain groups or facts (``\textit{There are risks that GenAI tools might generate biased and unfair visualisations if the training datasets are biased}'' - P4). Moreover, participants noted that a lack of diversity in training data can introduce inaccurate representation of multifaceted or culturally varied data into data comics (``\textit{If certain languages or racial groups predominate, the results will emphasise these characteristics over others}'' - P30). These comments suggested that participants might perceive the use of GenAI as reinforcing existing disparities in data comics, highlighting the need for diverse, well-curated datasets and carefully designed prompts to ensure equitable visualisations.

\textbf{\textit{Theme 3 - Threatening ownership and intellectual property.}} Participants (N=37) expressed concerns about the challenges surrounding the ownership and intellectual property rights of GenAI-generated images in data comics. A majority highlighted issues related to the unauthorised reproduction of copyrighted works. GenAI tools often directly ``\textit{make use of art or texts that were copyrighted}'' (P39), typically exploiting them ``\textit{without giving credit to their sources}'' (P45). Consequently, the ownership of these altered/reprocessed works remained ambiguous (``\textit{Does it belong to the AI company, the human prompter, or both}'' - P41), leading to a ``\textit{grey area in terms of profiting from AI-generated content}'' (P41). From participants’ perspectives, the integration of GenAI into data comics production was perceived as potentially complicating intellectual property considerations.

\textbf{\textit{Theme 4 - Reducing trustworthiness.}} This theme distilled referred to the limited trustworthiness of GenAI co-created content perceived by participants (N=23). The majority emphasised the need for critical oversight of content creation, arguing the outputs should be ``\textit{overseen by human}'' (P20). This content, generated by GenAI, was supposed to undergo scrutiny for ``\textit{credibility}'' (P39) and ``\textit{validity}'' (P55). P23 also highlighted the necessity of maintaining transparency about the role AI plays in the imagery production process. Moreover, a few participants shared their concerns regarding the infringement of data privacy, claiming there's a risk of sensitive information exposure by these visual artefacts co-created with GenAI (``\textit{Ensuring that personal data is anonymous and that the comics do not inadvertently reveal sensitive information is essential}'' - P37). These perspectives indicated that the integration of GenAI in producing data comics may diminish the credibility of the conveyed data insights due to the lack of trust in GenAI.

\section{Discussion}
\subsection{Summary of the Results and Research Questions}
Regarding \textbf{RQ1}, we found evidence about the superior effectiveness of data comics over conventional visualisations for interpreting data insights more accurately. This aligns with the findings from previous research that focused on data storytelling in general \citep{gomez2023personal,liang2024data,wang2019teaching,boucher2023educational}, embodying empirical evidence with significant results to reinforce the claims but now focused on data comics being effective for data insights communication to students.

For \textbf{RQ2}, we found strong evidence that data comics strengthen effective data insights communication across different question types, including information retrieval and comprehension (with sub-categories), compared to conventional visualisations. Similar to findings in \citet{shao2024data} on annotated charts, our results suggest that data comics, as a form of data storytelling, effectively guide audiences to insights and enhance accuracy even in tasks that require discerning multiple data points within a single visualisation.

Regarding \textbf{RQ3}, our findings indicate a moderate to strong relationship between visualisation literacy and accuracy rates for both visualisation conditions, supporting similar findings in previous studies \citep{shao2024data,milesi2025piecing}. Regardless of participants' visualisation literacy levels, accuracy rates were consistently higher with data comics, indicating that data comics are more effective at conveying insights. These results empirically support the widely held belief that data comics integrated with storytelling techniques benefit all levels of visualisation literacy \citep{figueiras2014narrative, maltese2015data}. Yet, the relationship between visualisation literacy and effectiveness appeared consistent across both conventional visualisations and data comics, meaning there is no equalising effect of data comics. In other words, participants with lower visualisation literacy still underperformed compared to their higher-literacy participants, regardless of the narrative and visual support provided by data comics.

Regarding \textbf{RQ4}, we confirmed participants' positive perceptions of data comics in promoting accessibility in data insights communication, consistent with indications in prior literature. Most participants (90\%) highlighted the combination of narrative elements and structured layouts, as described by \citet{bach2018design}, as making complex data more approachable. Our findings provide empirical evidence to support previous studies which have suggested the effectiveness of data comics in delivering nuanced information, particularly for novice audiences \citep{liang2024data, milesi2024s}. The perceived effectiveness in engagement echoes the findings from \citet{wang2019comparing}, \citet{gomez2023personal}, and \citet{milesi2024s}, who observed that data comics led to higher enjoyment and comprehension compared to conventional visualisations. Yet, this needs to be empirically tested in the context of data comics, especially since \citet{Zdanovic2022recall} did not find significant differences in information retrieval tasks between conventional visualisations and data storytelling visualisations.

Students also expressed concerns surrounding potential information overload, aligning with the findings by \citet{wang2019comparing} and \citet{Zdanovic2022recall}, stating that the textual and pictorial design can result in higher cognitive load. The issue of the challenge of generalisability, particularly in using data comics to represent complex data relationships, also emerged, echoing concerns in the literature about their broader applicability across different contexts (e.g., more serious or sensitive domains) \citep{chen2023does}. Moreover, the dependency on high-quality design emphasises the need for skilled or tailored design to effectively present comprehensive data insights through data comics, which is highlighted by \citet{wang2019comparing,Boucher2025instructional}.  

Finally, regarding \textbf{RQ5}, the results shed light on critical ethical issues regarding using GenAI technologies to assist in the creation of data comics. These are consistent with the broader ethical challenges growingly being identified in relation to the use of GenAI in educational contexts \citep{yan2024vizchat, farrelly2023generative}. Participants were notably worried about the inadvertent dissemination of inaccurate or misleading information, which is consistent with \citet{yan2024vizchat}'s findings. The risk of ``hallucination'' or generating nonfactual content, as noted by our participants, would potentially threaten the integrity of educational resources or systems created using GenAI. Furthermore, biases in GenAI outputs could perpetuate stereotypes and discrimination, an issue also highlighted by \citet{farrelly2023generative}. This reflects that the underlying training datasets, which often contain historical and algorithmic prejudices, can misrepresent facts or marginalise certain groups. 

Additionally, the challenge of intellectual property in GenAI applications, as discussed by \citet{chiu2023impact}, is echoed in our study, where participants expressed concerns about copyright issues in data comics co-created with GenAI. The ambiguity around authorship and ownership rights mirrors the broader debate in the literature about the implications of AI-generated content in educational contexts \citep{vincent2020trustworthy}, which often blurs the lines between original and derivative works. Moreover, the issue of diminished trustworthiness of GenAI outputs raised by our participants aligns with the broader call for transparency and human oversight in GenAI deployments, as emphasised in the literature \citep{holmes2022artificial}. The potential for GenAI to undermine equity and accessibility, particularly for marginalised groups or for learners with disabilities or limited language proficiency, also resonates with concerns discussed by \citet{farrelly2023generative} regarding the inclusiveness of AI technologies.

\subsection{Implications for HCI and Educational Technology Research}
Our findings advance HCI research by providing empirical evidence on how data storytelling approaches can enhance the communicative effectiveness of data. Prior work has demonstrated the potential of data storytelling to support information retrieval and comprehension, mainly in the form of annotated visualisations \citep{shao2024data, gomez2023personal, milesi2025piecing,Boy2015storytelling}. Yet, little is known about how different storytelling genres perform in practice. In particular, data comics, as a narrative visualisation genre \citep{segel2010narrative}, had not previously been evaluated in a comparative study against conventional visualisations. 
By focusing on students and the communicative power of data comics in educational contexts, our study offers empirical insights for HCI and educational technology researchers who are designing both general purpose and personalised learning interfaces. Such interfaces have the potential to transform data into engaging narratives that support comprehension, reflection, and learning. Integrating different forms of narrative visualisation techniques into educational interfaces can create new opportunities to examine how data storytelling fosters student engagement and understanding \citep{boucher2023educational}.

This contribution also speaks to ongoing HCI work on learning analytics interfaces and dashboards, particularly those designed to scaffold metacognition \citep{echeverria2025teamvision, martinez2020data,Zhang2025cvpis}. Our findings align with the broader shift in learning analytics, from being primarily visualisation-driven to becoming more pedagogically informed, focusing on supporting student learning through meaningful narratives rather than merely presenting charts \citep{paulsen2024LADs, fernandez2022beyond}.
Beyond the education domain, this work also opens new avenues for comparative research on data-driven storytelling modalities, highlighting the need to systematically evaluate the effectiveness of data comics, explanatory visualisations, and other narrative techniques across different learning and reflection contexts.

\subsection{Implications for Educational and Design Practitioners}

Our results demonstrate that data comics can significantly improve students' accuracy and comprehension, highlighting their potential as an effective pedagogical tool. For educational practitioners, such as teachers, data comics can be used to enrich lessons and assessments by making complex data more accessible and engaging. They can be especially valuable in subjects that demand data interpretation, such as science, social studies, and economics \citep{Kangtoonnote2021,vacca2022happen,kim2019datatoon,wang2019teaching,hasan2022playing}, where conventional visualisations often challenge even STEM students \citep{maltese2015data}. In K–12 contexts, visual storytelling can further foster engagement and provide early opportunities for learners to make sense of data \citep{milesi2024s}.
In this sense, GenAI offers a promising avenue to lower technical and creative barriers, enabling teachers and educational designers who may lack drawing or data analysis skills to create and integrate effective data comics into their lessons \citep{Boucher2025instructional}. 

From a constructivist perspective \citep{honebein1996seven,abderrahim2021theoretical}, data comics can encourage active sense-making of complex insights \citep{dangol2023constructionist}. The narrative format can help students connect new information to prior knowledge, while expressive imagery, sequences, and contextual scenarios create meaningful learning contexts. This alignment can support deeper cognitive processing, stronger mental models, reflective engagement, and improved long-term comprehension \citep{cashman2008power,phillips1995good}. Data comics can also be integrated into student-facing dashboards, supporting timely monitoring of progress and identification of learning gaps, while easing teachers’ interpretive workload \citep{fernandez2022beyond,cheng2024influence}.

For designers of educational interfaces, our results highlight both the promise and the limitations of data comics. The constrained space of this format can lead to oversimplification, making it unsuitable for certain types of data relationships. Therefore, the successful use of data comics requires careful design choices. Educators and designers must align comic formats with specific curriculum needs and learning outcomes; for instance, they may be effective in introductory visualisation courses \citep[e.g.,][]{wang2019teaching} but less suitable in advanced contexts. Moreover, like other data storytelling genres such as annotated visualisations \citep{shao2024data}, data comics may impose a particular perspective on the data. This raises issues of accountability in design decisions \citep{Kangtoonnote2021}, whether human- or algorithm-driven, and echoes concerns about bias in educational contexts \citep{milesi2024qualitative}.

The growing use of GenAI in creating educational materials (e.g., data comics) further amplifies these concerns \citep{maheshi2024data, binhammad2024investigating, linarespellicer2025breaking, milesi2024s}. Both researchers and educators must carefully address the ethical risks of misinformation and bias. For designers, this means integrating safeguards in the design phase of learning interfaces \citep{fernandez2022beyond, williamson2022review}. For educators, it involves critically selecting tools that uphold ethical standards. Only then can technology-augmented data storytelling approaches, such as data comics co-created with GenAI, provide students with learning experiences that are both engaging and trustworthy.

\subsection{Limitations and Future Work}
Our study has several acknowledged limitations in the stage of data collection and study design. In terms of our data collection, the sample size of 60 students may not fully represent the broader population that could benefit from data comics. These students, who were predominantly from similar academic backgrounds (i.e., STEM), may limit the generalisability of our findings to other groups with different professional contexts. Regarding the study design, by restricting our exploration to only four pairs of visualisations, the breadth of data comics and visualisation types assessed was limited, potentially affecting the applicability of our results across varied data scenarios. Future research could take into account involving a more diverse student group and extend the type and quantity of visualisations studied to promote a more comprehensive analysis \citep{pouliquen2004geographical}. 

Additionally, our study focused exclusively on the first two levels of Bloom's taxonomy, which constrained our evaluation to the retrieval of key data points and the comprehension of insights through data comics. Future research could expand upon this by including additional levels of the taxonomy, assessing more complex cognitive tasks such as the manipulation of data stories \citep{Arneson2018bloom, byrd2019using}.

Moreover, our study did not explicitly integrate pedagogical or learning design considerations into the creation or evaluation of data comics. While the comics were tested for effectiveness and efficiency in supporting tasks, they were not embedded within instructional activities or connected to broader learning outcomes. Future research should explore how data comics can be situated within educational contexts, such as teaching strategies, assessment designs, or reflective learning activities, to better align their use with pedagogical goals.

Lastly, while our approach leveraged GenAI to assist in the creation of data comics, it is worth noting that the procedures for creating the data comics still require prevalent human involvement. On the one hand, maintaining human oversight in the process is essential to ensure control, intentionality, and agency in how data insights are communicated \citep{ALFREDO2024100215}. 
Yet, future work could explore advanced techniques to streamline the creation of data comics that are trustworthy-for example, through neuro-symbolic architectures that embed strict logical constraints to mitigate hallucinations \citep{Shi2024constraint}, alongside the development of rigorous evaluation metrics to automatically assess the pedagogical validity of these artefacts \citep{Cui2025promises}.
%Yet, future work could explore advanced techniques to streamline the GenAI-assisted creation of data comics—for example, through more effective prompting strategies \citep{wang2024promptcharm} or multi-agent frameworks that scale different aspects of the design process \citep{ding2023designgpt}. 
However, such automation also amplifies ethical concerns, particularly around bias and accountability, since the storyteller, whether human or algorithmic, ultimately shapes which aspects of the data are highlighted.

\section{Concluding remarks}
This study highlights the potential of data comics as an effective tool for improving data insight comprehension among students with varied levels of visualisation literacy. By comparing conventional visualisations with data comics, we found that data comics enhance the identification of key information and data insights, which can be particularly advantageous for novice students. However, students' visualisation literacy remains a critical factor in determining how effectively they can use both conventional visualisations and data comics to comprehend insights. Despite their perceived benefits, data comics also present limitations, such as potential information overload and oversimplification, emphasising the need for careful balance between readability and detail. Additionally, while GenAI poses great potential to facilitate the creation process of data comics for pedagogical purposes, ethical considerations around misinformation, bias, and intellectual property in its applications highlight the importance of thoughtful implementation and oversight in educational contexts.

%\backmatter
% \section*{Author contributions}

% This is an author contribution text. This is an author contribution text. This is an author contribution text. This is an author contribution text. This is an author contribution text.
%TC:ignore
\section*{Acknowledgements}
We acknowledge the adoption of a large language model-based tool (OpenAI's ChatGPT) in assisting the creation of pictorial elements for our research purposes. In addition, ChatGPT was used to support text refinement and to improve clarity for our manuscript. All content produced with the assistance of LLM was reviewed, edited and verified by the authors, who take full responsibility for the final text.

% \section*{Financial disclosure}

% None reported.

\section*{Conflict of interest}

The authors declare no potential conflict of interests.

\section*{Data Availability Statement}
The data that support the findings of this study are available from the corresponding author upon reasonable request.

\bibliographystyle{unsrtnat}
\bibliography{reference}

\appendix

\section{Demographic Questions}\label{appendix:a1_demographic}

\begin{enumerate}

\item What gender do you identify yourself as?
\begin{itemize}
    \item Male
    \item Female
    \item Non-binary / gender diverse
    \item Other: Please type in your answer
    \item Prefer not to say
\end{itemize}

\item What is the highest degree or level of education you have completed?
\begin{itemize}
    \item High school graduate
    \item Bachelor's Degree
    \item Master's Degree
    \item Ph.D.
    \item Other: Please type in your answer
    \item Prefer not to say
\end{itemize}

\item What is your major or primary field of study?
\begin{itemize}
    \item STEM (Science, Technology, Engineering, Mathematics)
    \item Business or Economics
    \item Arts or Humanities
    \item Social Sciences
    \item Other: \underline{\hspace{3cm}}
\end{itemize}

\item Choose all the circumstance that describes your experience with data visualisation:
\begin{itemize}
    \item I never heard of it
    \item I only heard of it but don’t know much
    \item I have encountered data visualisations on newspapers/apps/websites
    \item My job requires me to read data visualisations
    \item I'm a professional in this area and create them myself
\end{itemize}

\end{enumerate}
%-------------------------------------------------

\section{Mini-Visualisation Literacy Assessment Test}\label{appendix:a2_minivlat}

In our research, we adopted \href{https://doi.org/10.1111/cgf.14809}{Mini-VLAT} \citep{pandey2023mini} to test participants' visualisation literacy. It is a validated visualisation literacy test containing 12 multiple-choice questions about twelve different types of charts. For each question, participants were given 25 seconds to decide an answer. If the allocated time was exceeded, the survey automatically advanced to the next question.

If participants were unsure about an answer, they could select ``Skip'' and continue to the next question. Participants were encouraged to answer to the best of their ability. Visualisations may take some time to load.

%-------------------------------------------------

\section{Comparative Visualisation Questions}\label{appendix:a3_comparative}

The following set of questions contains 16 multiple-choice questions about four different data visualisations received by the participants during the study:

\noindent For each question, you will be given a visualisation. After attempting to answer the questions, if you are unsure about the answer, you may select 'Skip' and continue to the next question.

\noindent However, please do try your BEST to answer the questions as skipping THREE questions may result in your response being invalid and excluded from the research study.

\subsection{Comparative Pair 1 (CV1 / DC1)}

\begin{figure}[!htbp]
    \centering
    \includegraphics[width=\linewidth]{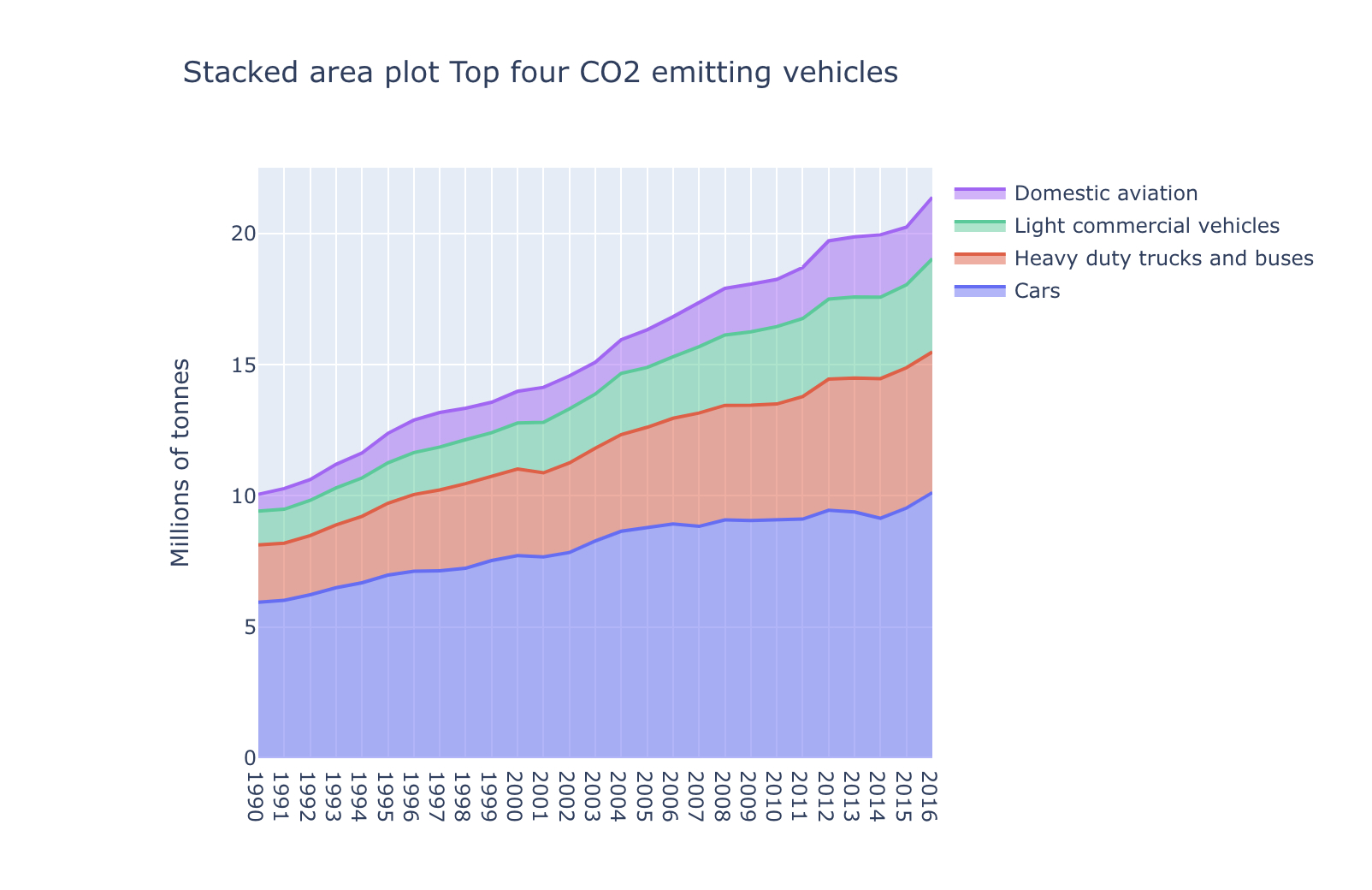}
    \caption{Comparative Visualisations 1: Conventional Visualisation}
    \label{fig:a-cv1}
\end{figure}
\begin{figure}[!htbp]
    \centering
    \includegraphics[width=0.55\linewidth]{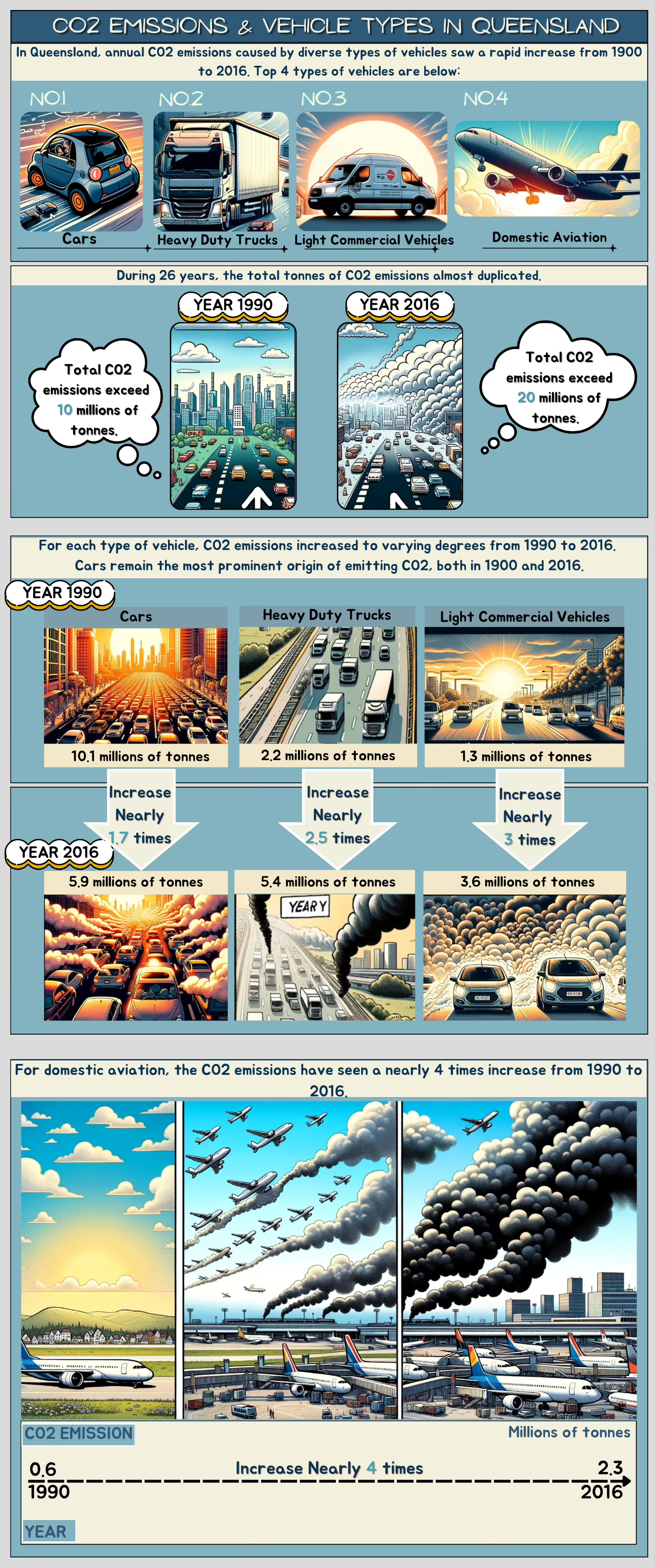}
    \caption{Comparative Visualisations 1: Data Comics}
    \label{fig:a-dc1}
\end{figure}

\subsubsection{Information Retrieval questions}

\begin{enumerate}
\item Which number is the closest to the amount of total CO$_2$ emissions in 2016?
\begin{enumerate}
    \item 10M tonnes
    \item 15M tonnes
    \item 25M tonnes
    \item 30M tonnes
\end{enumerate}

\item Which type of vehicles is the most prominent source of CO$_2$ emissions?
\begin{enumerate}
    \item Domestic aviation
    \item Light commercial vehicles
    \item Heavy duty trucks and buses
    \item Cars
\end{enumerate}
\end{enumerate}

\subsubsection{Comprehension questions}

\begin{enumerate}
\item Around how many times did light commercial vehicles CO$_2$ emissions increase from 1990 to 2016?
\begin{enumerate}
    \item 2.0
    \item 2.5
    \item 3.0
    \item 3.5
\end{enumerate}

\item Choose the incorrect statement:
\begin{enumerate}
    \item CO$_2$ emissions resulting from domestic aviation have rapidly increased from 1990 to 2016
    \item CO$_2$ emissions from heavy-duty trucks and buses increased nearly 2.5 times
    \item CO$_2$ emissions from domestic vehicles almost quadrupled
    \item The total tonnes of CO$_2$ emissions almost tripled
\end{enumerate}
\end{enumerate}
%-------------------------------------------------

\subsection{Comparative Pair 2 (CV2 / DC2)}

\begin{figure}[!htbp]
    \centering
    \includegraphics[width=\linewidth]{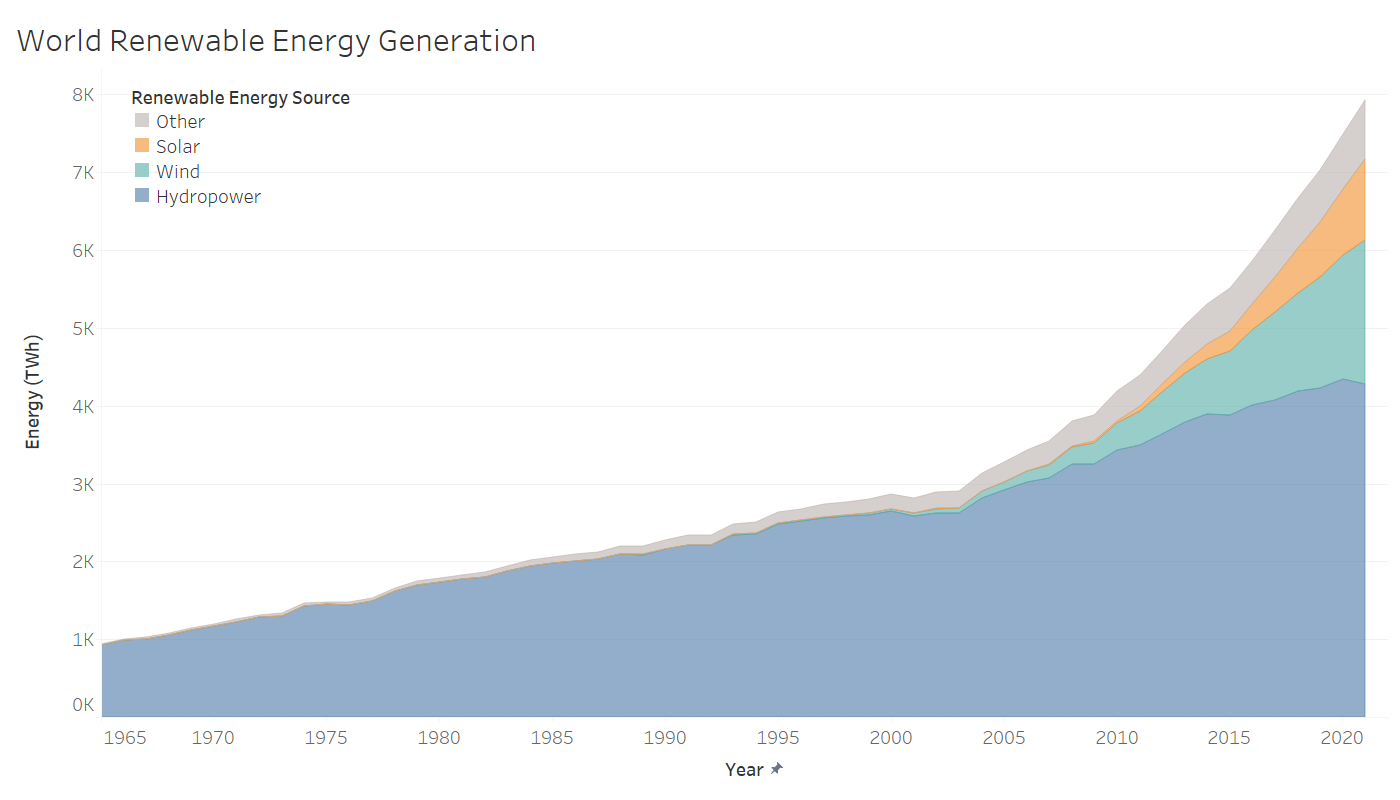}
    \caption{Comparative Visualisations 2: Conventional Visualisation}
    \label{fig:a-cv2}
\end{figure}
\begin{figure}[!htbp]
    \centering
    \includegraphics[width=0.65\linewidth]{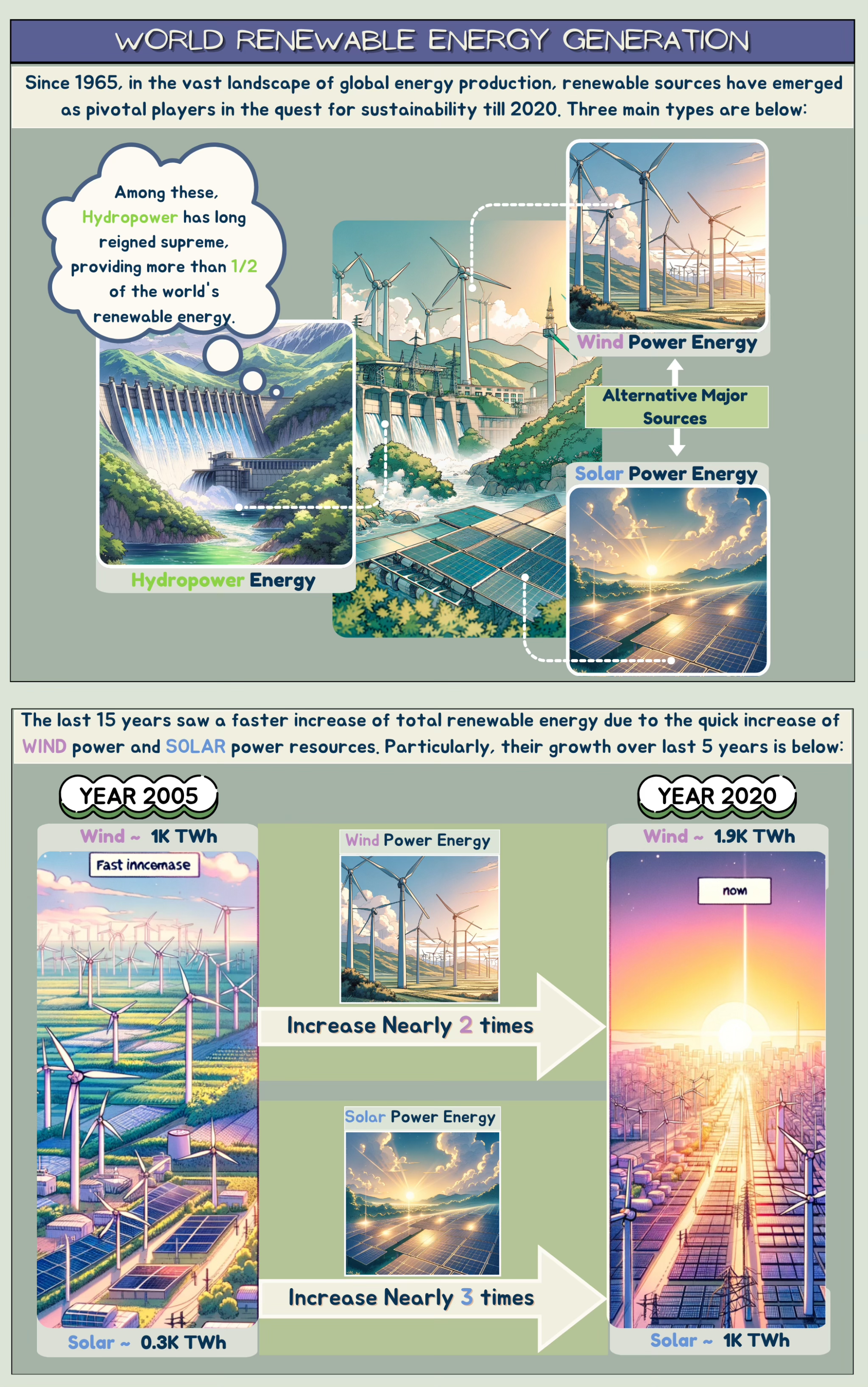}
    \caption{Comparative Visualisations 2: Data Comics}
    \label{fig:a-dc2}
\end{figure}

\subsubsection{Information Retrieval questions}
\begin{enumerate}
\item Which renewable energy source dominated among all types of sources?
\begin{enumerate}
    \item Solar
    \item Wind
    \item Hydropower
    \item Others
\end{enumerate}

\item Since which year did the total renewable energy see a faster increase?
\begin{enumerate}
    \item 1965
    \item 2000
    \item 2005
    \item 2020
\end{enumerate}
\end{enumerate}

\subsubsection{Comprehension questions}
\begin{enumerate}
\item Around how many times did wind and solar power energy resources increase from 2015 to 2020?
\begin{enumerate}
    \item 1.5
    \item 2.0
    \item 2.5
    \item 3.0
\end{enumerate}

\item Choose the incorrect statement:
\begin{enumerate}
    \item Wind power resources increased faster than solar power resources from 2015 to 2000
    \item Solar power resources have been rapidly increasing since 2005
    \item More than half of renewable energy comes from hydropower
    \item Wind and solar power resources are becoming alternative major sources
\end{enumerate}
\end{enumerate}
%-------------------------------------------------

\subsection{Comparative Pair 3 (CV3 / DC3)}

\begin{figure}[!htbp]
    \centering
    \includegraphics[width=\linewidth]{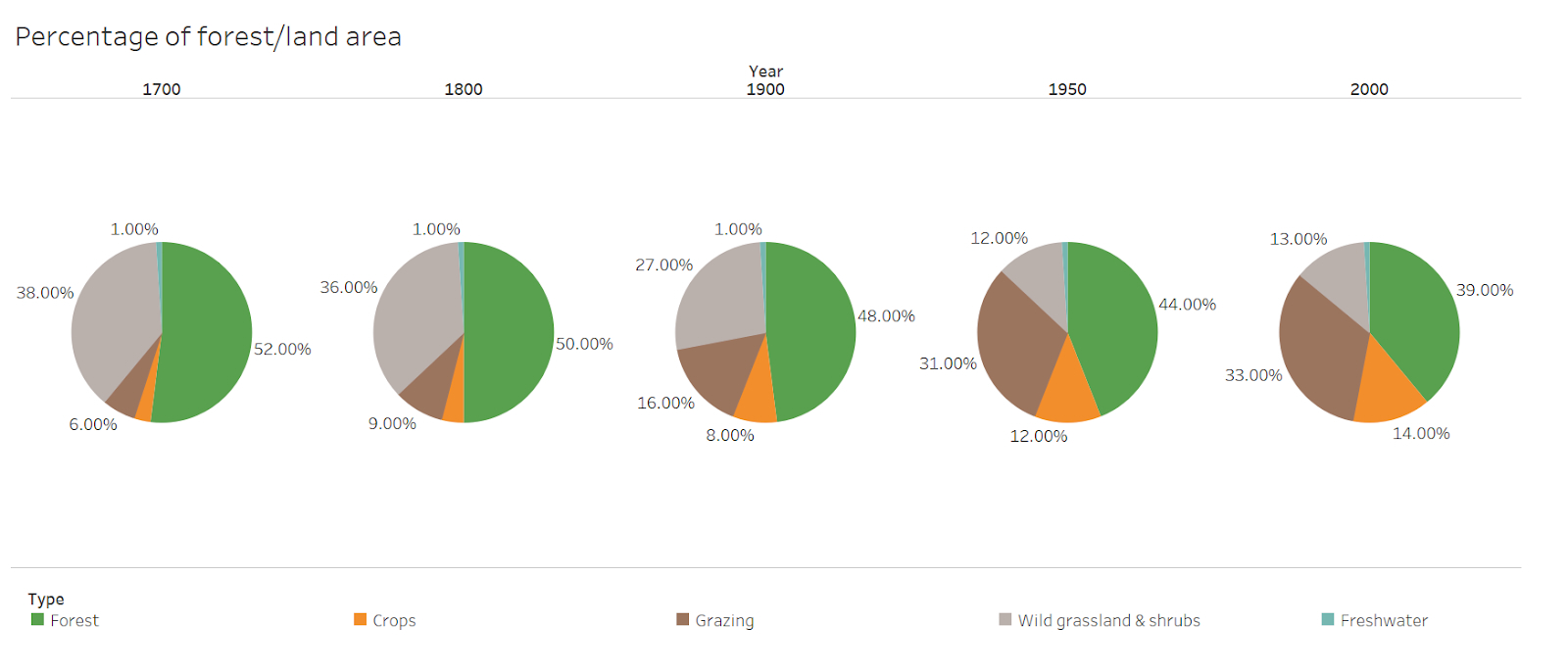}
    \caption{Comparative Visualisations 3: Conventional Visualisation}
    \label{fig:a-cv3}
\end{figure}
\begin{figure}[!htbp]
    \centering
    \includegraphics[width=0.55\linewidth]{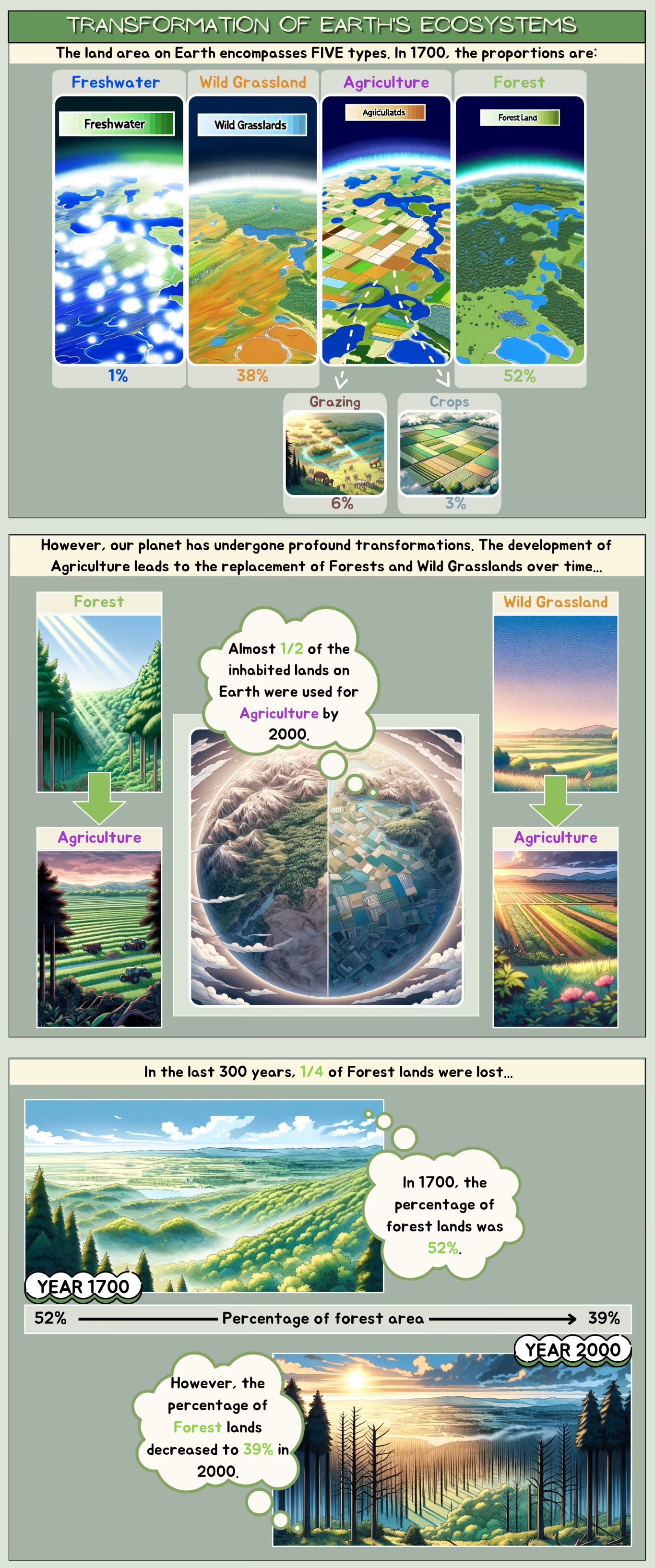}
    \caption{Comparative Visualisations 3: Data Comics}
    \label{fig:a-dc3}
\end{figure}

\subsubsection{Information Retrieval questions}

\begin{enumerate}
\item What proportion of land was used for Grazing in 1700?
\begin{enumerate}
    \item 1\%
    \item 3\%
    \item 6\%
    \item 8\%
\end{enumerate}

\item What is the difference in proportion of Wild Grassland \& Shrubs and Forest in 1700?
\begin{enumerate}
    \item 28\%
    \item 25\%
    \item 23\%
    \item 14\%
\end{enumerate}
\end{enumerate}

\subsubsection{Comprehension questions}

\begin{enumerate}
\item What may be the percentage of Agriculture lands by 2000?
\begin{enumerate}
    \item 5\%
    \item 9\%
    \item 22\%
    \item 47\%
\end{enumerate}

\item Choose the incorrect statement: 
\begin{enumerate}
    \item 1/4 of Forest was lost within 300 years
    \item Almost half of the inhabited lands on Earth were used for Agriculture by 2000
    \item The development of Agriculture leads to the replacement of Forests and Wild Grasslands
    \item Forest lands decreased because of the increase in the area of Wild Grassland \& Shrubs
\end{enumerate}
\end{enumerate}
%-------------------------------------------------

\subsection{Comparative Pair 4 (CV4 / DC4)}

\begin{figure}[!htbp]
    \centering
    \includegraphics[width=\linewidth]{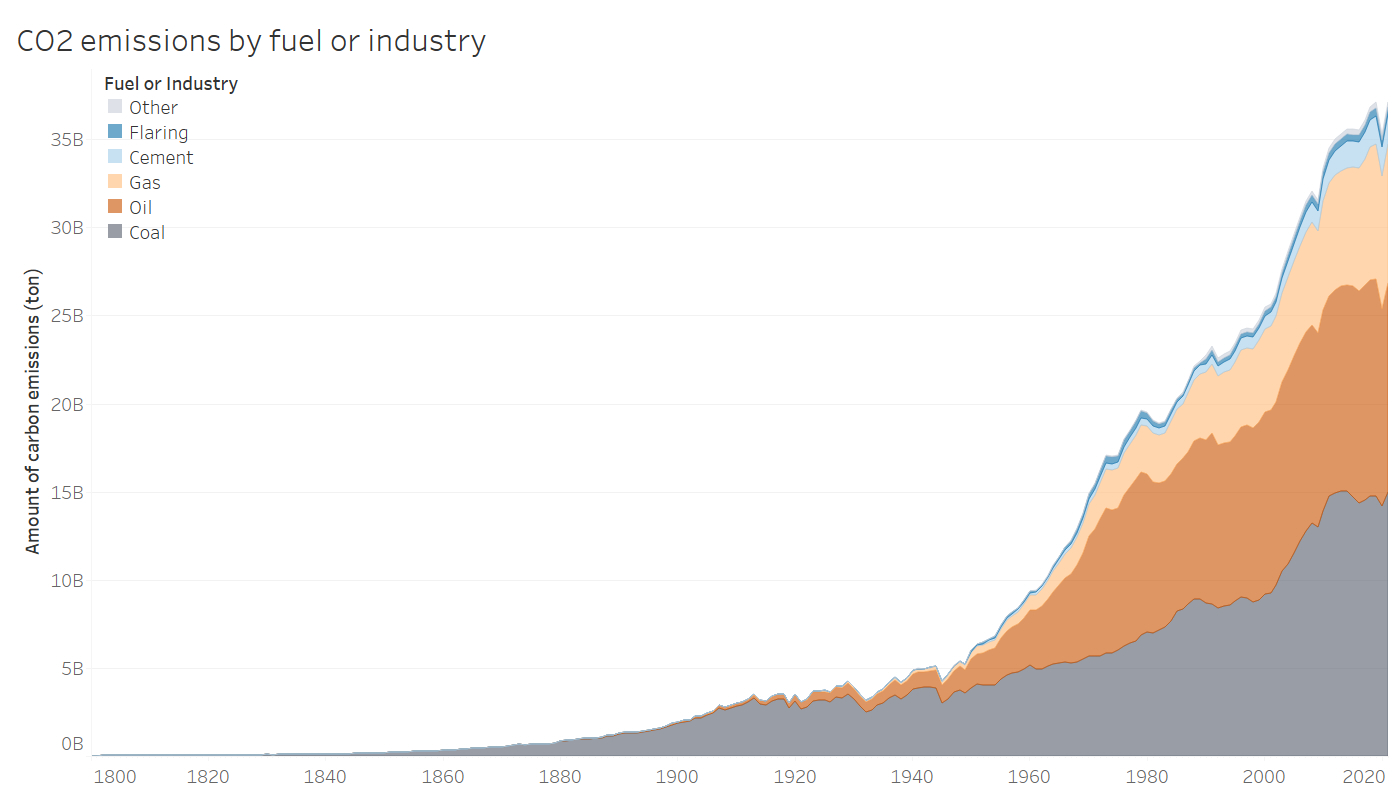}
    \caption{Comparative Visualisations 4: Conventional Visualisation}
    \label{fig:a-cv4}
\end{figure}
\begin{figure}[!htbp]
    \centering
    \includegraphics[width=0.55\linewidth]{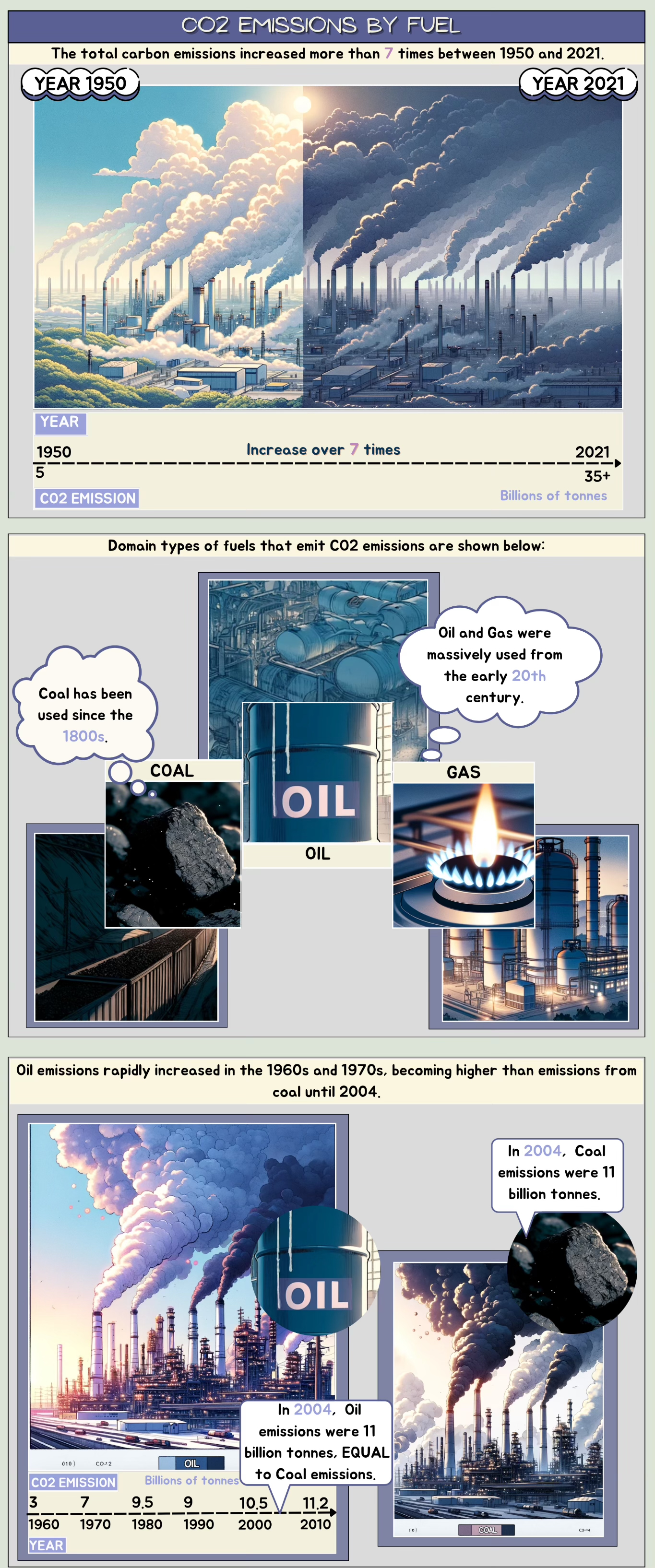}
    \caption{Comparative Visualisations 4: Data Comics}
    \label{fig:a-dc4}
\end{figure}

\subsubsection{Information Retrieval questions}

\begin{enumerate}
\item Around how many billion tons of CO$_2$ were emitted in 2021? 
\begin{enumerate}
    \item 5B tons
    \item 20B tons
    \item 25B tons
    \item 35B tons
    \item I am not sure about the answer
\end{enumerate}

\item Around how many billion tons of CO$_2$ were emitted from oil in 1990?
\begin{enumerate}
    \item 8B tons
    \item 9B tons
    \item 10B tons
    \item 15B tons
\end{enumerate}
\end{enumerate}

\subsubsection{Comprehension questions}

\begin{enumerate}
\item Around how many times did total CO$_2$ emissions increase from 1950 to 2021? 
\begin{enumerate}
    \item 5
    \item 6
    \item 7
    \item 8
\end{enumerate}

\item Please choose the sentence that is incorrect: 
\begin{enumerate}
    \item Oil emissions equal to emissions from coal in 2004
    \item Oil and gas were massively used from the early 21$^{st}$ century
    \item Coal has been used since the 1800s
    \item Oil emissions increased rapidly before 1980s
\end{enumerate}
\end{enumerate}
%-------------------------------------------------

\section{Open-ended Perception Questions}\label{appendix:a4_perception}

This section consists of open-ended questions aimed at understanding the specific attributes that may enhance or impede the communication of data insights through data comics. Please share detailed responses based on your observations and experiences with the materials provided, focusing on features, instances of effectiveness, and any perceived limitations.

\begin{enumerate}

    \item What specific features of data comics contribute to their effectiveness in \textbf{communicating data insights} compared to conventional visualisations? Elaborate on how the feature enhances the effectiveness.

    \item What specific features of data comics impact their effectiveness in \textbf{engaging audiences} compared to conventional visualisations? Elaborate on how the feature enhances the effectiveness.

    \item Can you provide examples of instances where you believe data comics have excelled in communicating complex data insights more effectively than conventional visualisations? What made these instances particularly effective?

    \item Conversely, are there any limitations or drawbacks of data comics that you perceive as hindering their effectiveness in communicating data insights compared to conventional visualisations? If so, what are they?
    
\end{enumerate}

%-------------------------------------------------

\section{Ethical Implication Questions}\label{appendix:a5_ethical}

The images you've seen in the data comics were developed using GenAI technology, a method that holds promise for transforming how we visualise and interpret data.

In this section, we seek your feedback on potential ethical aspects of using GenAI-powered data comics. 
\begin{enumerate}

\item Do you believe there are concerns about potential \textbf{misinformation} associated with the use of GenAI-powered data comics?
\begin{enumerate}
    \item Yes
    \item No
\end{enumerate}

\item Do you believe there are concerns about potential \textbf{bias and fairness} associated with the use of GenAI-powered data comics?
\begin{enumerate}
    \item Yes
    \item No
\end{enumerate}

\item Do you believe there are concerns about potential \textbf{ownership and intellectual property} associated with the use of GenAI-powered data comics?
\begin{enumerate}
    \item Yes
    \item No
\end{enumerate}

\item If any of your answers to the previous questions is Yes, please elaborate your concerns: \underline{\hspace{3cm}}

\item What other ethical concerns, if any, do you have regarding the use of GenAI-powered data comics in communicating data insights?

\end{enumerate}

%-------------------------------------------------

\section{Content Relation Categories for Data Insights}\label{appendix:a6_content}

We classified each insight into one or more categories following the framework proposed by \citet{bach2018design}, specifically the six content relation categories that scaffolded the selection of panel layouts to form a logical narrative:
\begin{enumerate}
    \item \textbf{Narrative} (i.e., insights that introduced a problem, question, or context to engage the reader and set up the story);
    \item \textbf{Temporal} (i.e., insights that highlighted changes over time, such as trends, progressions, or historical comparisons);
    \item \textbf{Faceting} (i.e., insights that compared multiple subjects, perspectives, or scenarios within the dataset);
    \item \textbf{Visual Encoding} (i.e., insights that explain how a visualisation works, guiding the audience through its interpretation);
    \item \textbf{Granular} (i.e., insights that zoomed into specific data points from the dataset for in-depth examination);
    \item \textbf{Spatial} (i.e., insights that related to geographic positioning, movement, or spatial relationships within the data).
\end{enumerate}

\section{Participant Demographics}\label{appendix:a7_demographics}
The participants identified themselves as male (27), female (32), and non-binary (1), with an average age of 27 years old. All were current students and most were studying towards a Bachelor's degree (11) and Master's degree (27), while others towards a doctoral degree (22). With regard to their majors, most of them were from STEM (43), while others primarily studied in Business or Economics (9), Arts or Humanities (2), Social Science (2), Digital Communication (1), Environment (1), Health Sciences (1), and English Philology (1). The participants came from six different regions, with the majority coming from Oceania (32), Africa (11), Europe (7) and North America (6). The remainder were from South America (3) and Asia (1).

\section{Prompting Guidelines for Generating Data Comic Elements}\label{image creation}
While DALL-E 3 is capable of generating high-quality images, it falls short in consistently maintaining the aesthetic visual styles \citep{albaghajati2023exploring} and can struggle to interpret convoluted prompts \citep{liu2022design,feng2023promptmagician}. This typically leads to outputs that do not fully align with the narrative or break the flow of a comic strip, making it challenging for an audience to engage with the story as intended.  
%Therefore, drawing inspiration from the work by \citet{li2023can}, \citet{liu2022design}, and \citet{oppenlaender2023taxonomy}, we developed several prompting guidelines tailored for generating visual elements used in crafting data comics effectively and coherently:
Therefore, we adopted a set of pragmatic prompting guidelines, adapted from prior work \citep{li2023can, liu2022design, oppenlaender2023taxonomy}, to reliably and effectively generate the visual elements for our data comics. Below, we report these guidelines to promote reproducibility:
 
\begin{itemize}
    \item \textbf{Context-based prompting:} 
 
     \texttt{Prompt = [context information] + [keywords description] + \newline [additional information] + [style]}
\end{itemize} 
 
This guideline integrates context about the topic or background, a clear description of the main requirements, any extra details tailored for special preferences, and a specified artistic style. Its effectiveness was especially notable in scenarios where the generated images were not only visually appealing but also contextually appropriate. An example prompt: \textit{We are creating a comic for the topic of ``CO2 emissions from vehicle'' (context information). Can you create an image presenting ``cars'' (keywords information), no characters (additional information), using high saturation colours, comic book art style and soft light effect (style)?} One example of the output from this prompt was the car image, as illustrated in Figure~\ref{fig:example panel}.
 
%(See the corresponding car image from Figure~\ref{fig:example panel} for the output from this prompt).
\begin{itemize}
    \item \textbf{Example-based prompting:} 
    
    \texttt{Prompt = [context information] + [keywords description] + \newline [additional information] + [example style]}
\end{itemize}
 
This guideline started with uploading a sample image to set a visual baseline. Similar to context-based prompting, a topic description was provided along with context information, enriched with additional details needed to refine the outcome. Distinguished from the context-based prompting guideline, the [style] was stabilised by prompting a command such as ``the style of comic should be aligned with the uploaded example image'' to ensure the alignment of visual tone. This was particularly suitable for maintaining a consistent style across multiple targeted images. An example prompt: \textit{We are creating a comic for the topic of ``CO2 emissions from vehicle'' (context information). Can you create an image presenting ``domestic aviation'' (keywords information), no characters (additional information)? The style of the comic should be aligned with the uploaded example image\footnote{The car image was used as the example.} (example style).} One example of the output from this prompt was the domestic aviation image, as illustrated in Figure~\ref{fig:example panel}.
 
%(See the corresponding domestic aviation image from Figure~\ref{fig:example panel} for the output from this prompt)
 
\begin{itemize}
    \item \textbf{Style-alternation iterative prompting.}
\end{itemize}
This guideline involved iteratively modifying one or more styles of the prompt (e.g., high saturation colours) and re-prompting to refine the image according to specific requirements. By adjusting these keywords incrementally, the generative model explored various visual possibilities, gradually converging towards the desired image. Iterative prompting allowed for progressive adjustments and nuanced fine-tuning processes until the final image closely aligned with the intended outcome. An example prompt: \textit{Can you revise the generated image by using high saturation colours?}
 
\begin{itemize}
    \item \textbf{Self-evaluative iterative prompting.}
\end{itemize}
 
This guideline referred to engaging GenAI such as ChatGPT-4 in self-evaluating processes, thereby refining the initially generated response automatically. This approach harnessed GenAI's capacity for self-awareness to enhance the quality of generated visuals, producing more precise and tailored outcomes. An example prompt: \textit{Can you evaluate the previous prompt for generating the comic and identify areas of improvement?} 
 
%Guiding by these prompting guidelines, we were able to get desired images such as the ones illustrating each type of vehicles (e.g., cars, heavy duty tracks) showcased in Figure \ref{fig:example panel} and the figures illustrating the amount of CO2 emissions both in year 1990 and year 2016 shown in Figure \ref{fig:contrast pattern}.

\section{Design Process for Crafting Data Comics}\label{sec:datacomicprocess}
As suggested by \citet{bach2018design}, depicting a data comic story involves two key dimensions: content relation and layout. To address these, the authors proposed a set of design patterns for data comics that capture how information can be structured visually and how narration can be effectively crafted \citep{bach2018design}. These patterns enable narrative elements to be organised into an appropriate sequence, with the aim of facilitating coherent narrative construction.
Following their proposed design patterns, we developed a two-stage process for crafting data comics. This process separates the human-led narrative design from the human–AI co-creation of visual elements, clarifying where generative AI contributed to the workflow.
 
\paragraph*{\textbf{Stage 1: Human-led narrative design.}}
 The first stage focused on establishing the narrative structure of the data comic. This involved three main steps:
 
\begin{enumerate}
[label*=\arabic*.]
 \item \textbf{Extracting insights from data.}
 Key insights are identified from the dataset to be communicated, guided by principles of data storytelling. This step is fully human-driven, ensuring that the selected insights align with audience needs and the intended communicative purpose \citep{knaflic2015storytelling,dykes2015data}.
 
 \noindent\textit{Example:} From a dataset on global emissions, one extracted insight was: \textit{“the total tonnes of CO2 emissions almost doubled from 1990 to 2016.”}
 
\item \textbf{Classifying insights into content relation categories.}
 Each insight is then classified into one or more of the six content relation categories defined by \citet{bach2018design} (Narrative, Temporal, Faceting, Visual Encoding, Granular, and Spatial; See Appendix \ref{appendix:a6_content} for descriptions of the six categories). This categorisation scaffolds the spatial arrangement of panels and organisation of insights on the page into coherent story elements to guide the reader's attention effectively.
 
 \noindent\textit{Example:} The CO2 emissions insight was classified as both \textbf{Temporal} (since it highlighted change over time) and \textbf{Faceting} (since it compared two time points). Another insight, \textit{“wind and solar power resources are rapidly increasing as alternative sources of renewable energy compared to Hydropower,”} was categorised as \textbf{Narrative} (introducing context) and \textbf{Granular} (focusing on specific data points).
 
\item \textbf{Mapping insights to panel layouts.}
 Based on the assigned categories, insights are mapped to panel layouts (e.g., Parallel, Annotated) to determine the sequencing and spatial arrangement of panels. This step ensures that the logical flow of the story is preserved prior to visual production.
 
 \noindent\textit{Example:} For the CO2 emissions insight (Temporal + Faceting), a \textbf{Parallel} layout was selected to show panels side by side for direct comparison. For the renewable energy insight (Narrative + Granular), an \textbf{Annotated} layout was chosen to allow smaller panels to highlight contextual details within a larger panel.
 
 \end{enumerate}

\begin{figure}[htbp!]
    \centering
    \includegraphics[width=0.8\linewidth]{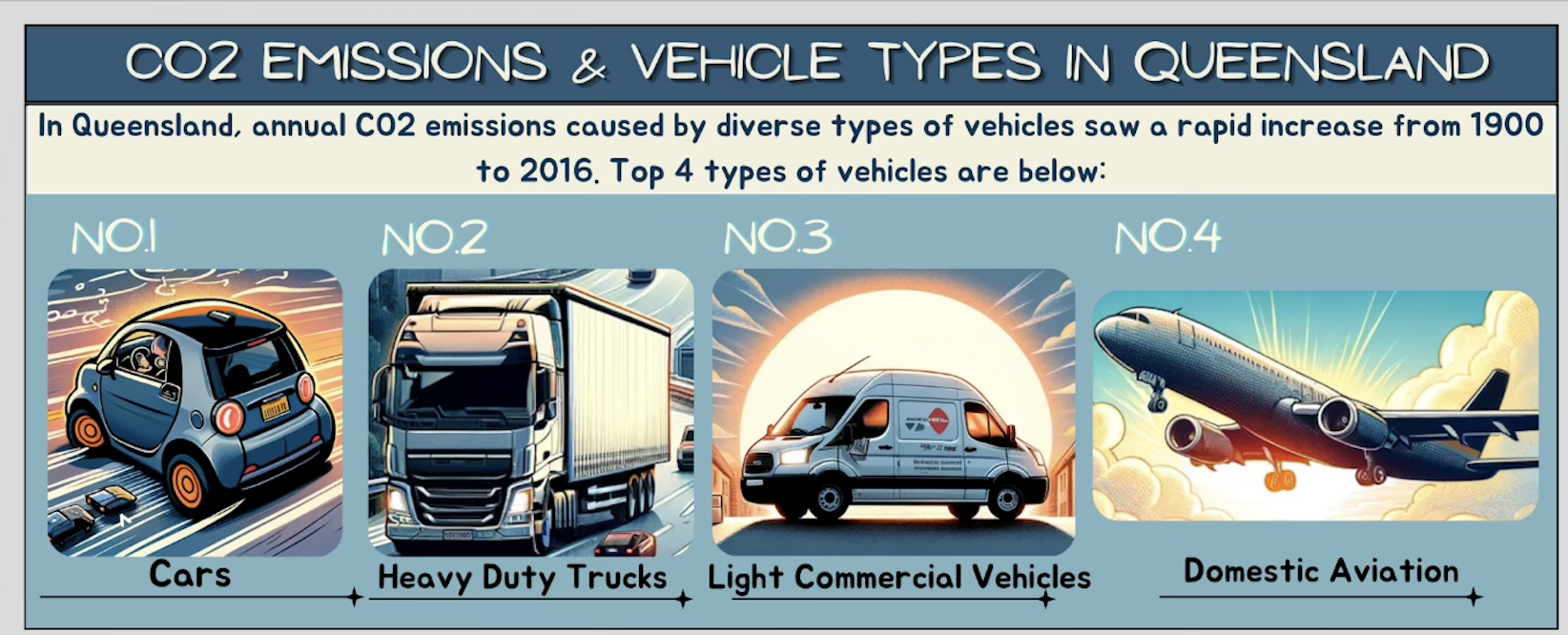}
    \caption{Example snapshot of data comics illustrating the \textit{exposé} pattern about CO2 emission.}
    \label{fig:example panel}
\end{figure}
 
\paragraph*{\textbf{Stage 2: Human–AI co-creation of comic strips.}}
 Once the narrative structure is defined, the second stage focuses on producing the comic strip through an iterative collaboration between humans and generative AI tools. This involves three further steps:

 \begin{figure}[htbp!]
        \centering
        \includegraphics[width=0.8\linewidth]{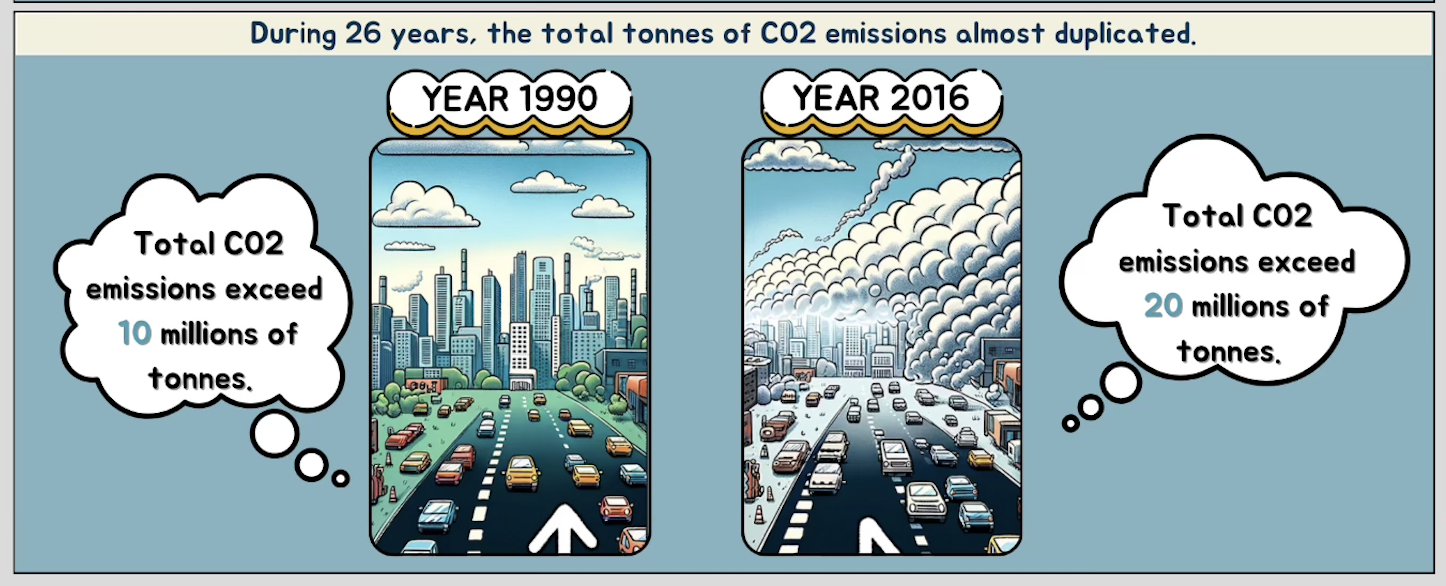}
        \caption{Example snapshot of data comics illustrating the \textit{contrast} pattern.}
        \label{fig:contrast pattern}
\end{figure}
 
\begin{enumerate}[label*=\arabic*.]
 \setcounter{enumi}{3}
 \item \textbf{Generating and refining AI images.}
 The guidelines described in Section \ref{image creation} are employed to produce candidate illustrations based on human-crafted prompts that reflect the identified insights (e.g., objects, characters, or locations). Humans iteratively refine these prompts, curate outputs, and select images that best align with the intended narrative. 
 
 \noindent\textit{Example:} The context-based prompting guideline (Section \ref{image creation}) was used to create an initial image for a comic strip, and the style-alternation iterative prompting approach was adopted to continuously refine the keywords in the context-based prompting guideline until a desired image style was achieved (e.g., the car image showcased in Figure \ref{fig:example panel}). Following this, the example-based prompting guideline was used to produce relevant figures (e.g., the heavy-duty trucks, light commercial vehicles and domestic aviation images in Figure \ref{fig:example panel}), using the initial image (e.g., the car image) as the example.  
 %guidelines x,y (Section \ref{image creation}) were used to get desired images such as the ones illustrating each type of vehicle (e.g., cars, heavy-duty trucks) showcased in Figure \ref{fig:example panel}. guidelines x, y, z (Section \ref{image creation}) were used to generate the figures illustrating the amount of CO2 emissions both in the year 1990 and year 2016, shown in Figure \ref{fig:contrast pattern}.
 
\item \textbf{Building comic strips.}
 The curated AI-generated images are integrated into the predefined panel layouts. Human designers ensure coherence between visuals and narrative flow, balancing stylistic consistency with the communicative goals of the comic.
 
 \noindent\textit{Example:} The four vehicle images were placed in a grid layout, each annotated as ``No.1'' through ``No.4'', supporting exposé pattern (see Figure~\ref{fig:example panel}). The cityscapes were arranged in parallel panels labelled ``Year 1990'' and ``Year 2016'' to reinforce the temporal compare/contrast pattern (see Figure~\ref{fig:contrast pattern}).   
 
\item \textbf{Adding storytelling elements.}
 Finally, storytelling features are layered onto the panels. Titles and subtitles help distil the dataset topics \citep{knaflic2015storytelling}, annotations (e.g., labels, arrows, speech balloons) support clarification of insights \citep{guerin2017ontology}, and highlights (e.g., bold text, colour contrasts) emphasise key statistics and trends \citep{ware2019information}.
 
\noindent\textit{Example:} In the vehicle breakdown strip, subtitles contextualised the scope (``Top 4 types of vehicles in Queensland''), and labels emphasised the categories (Cars, Heavy Duty Trucks, etc.; See Figure~\ref{fig:example panel}). Figure~\ref{fig:contrast pattern}, the CO2 comparison strip, shows the use of speech balloons to highlight the totals (``10 million tonnes'' vs. ``20 million tonnes''). 
 
\end{enumerate}

% Additional supporting information may be found in the
% online version of the article at the publisher’s website.

%\nocite{*}% Show all bib entries - both cited and uncited; comment this line to view only cited bib entries;

% \section*{Author Biography}

% \begin{biography}{\includegraphics[width=76pt,height=76pt,draft]{empty}}{
% {\textbf{Author Name.} Please check with the journal's author guidelines whether
% author biographies are required. They are usually only included for
% review-type articles, and typically require photos and brief
% biographies for each author.}}
% \end{biography}

%TC:endignore
\end{document}